\documentclass[twocolumn,twosides,letterpaper,10pt]{IEEEtran}
\usepackage{cite,url,color}
\usepackage{graphicx} %[pdftex]
\graphicspath{{./pdf/}{./jpeg/}{./figs/}{./eps/}}
\DeclareGraphicsExtensions{.pdf,.jpeg,.png,.eps}
\usepackage{multirow}
\usepackage{booktabs}
\usepackage[cmex10]{amsmath}
\usepackage{amssymb,bm}
\usepackage{algpseudocode} % we use this
\newcommand{\algrule}[1][.5pt]{\par\vskip.25\baselineskip\hrule height #1\par\vskip.25\baselineskip}
\usepackage[caption=false,font=footnotesize]{subfig}
\usepackage{hyperref}
\hypersetup{
  colorlinks=true,
  linkcolor=black,
  citecolor=black,
  filecolor=black,
  urlcolor=black,
  pdfborder={0 0 0}
}
\usepackage[amsmath,thmmarks]{ntheorem}

\newcommand{\ve}[1]{\mathbf{{#1}}}
\newcommand{\abs}[1]{\ensuremath{\vert #1\vert}}
\newcommand{\norm}[1]{\ensuremath{\Vert #1\Vert}}

\begin{document}

\title{Bayesian De-quantization and Data Compression for Low-Energy Physiological Signal Telemonitoring}
% author names and IEEE memberships, place (~) in author names to prevent breaking
\author{Benyuan~Liu$^{*}$,~\textit{Student Member,~IEEE},~Hongqi~Fan,~and~Qiang~Fu,~Zhilin~Zhang,~\textit{Member,~IEEE}%
\thanks{\textit{Asterisk indicates corresponding author.}}%
\thanks{$^*$B. Liu is with The Science and Technology on Automatic Target Recognition Laboratory, %
National University of Defense Technology, Changsha, Hunan 410074, China and also with %
the Department of Biomedical Engineering, Fourth Military Medical University, Xi'an 710032, China. (e-mail: byliu@fmmu.edu.cn)}
\thanks{H.-Q. Fan, Q. Fu are with The Science and Technology on Automatic Target Recognition Laboratory, %
National University of Defense Technology, Changsha, Hunan, 410074, China (e-mail: \{fanhongqi,fq\}@nudt.edu.cn)}
\thanks{Z. Zhang is with The Emerging Technology Lab, %
Samsung Research America - Dallas, 1301 East Lookout Drive, %
Richardson, TX 75082, USA. (e-mail: zhilinzhang@ieee.org)}%
\thanks{Manuscript received \today{}.}%
}

\markboth{Submitted to IEEE Journal of Biomedical and Health Informatics~Vol.~X, No.~X, 2015}%
{Liu \MakeLowercase{\textit{et al.}}: Block Sparse De-Quantize}



\maketitle

%-------------------------------------------------------------------------
\begin{abstract}
\boldmath
%% quick and cut!
We address the issue of applying quantized compressed sensing (CS) on low-energy telemonitoring.
So far, few works studied this problem in applications where signals were only approximately sparse.
We propose a two-stage data compressor based on quantized CS,
where signals are compressed by compressed sensing and then the compressed measurements are quantized with only $2$ bits per measurement.
This compressor can greatly reduce the transmission bit-budget.
To recover signals from underdetermined, quantized measurements, we develop a Bayesian De-quantization algorithm.
It can exploit both the model of quantization errors and the correlated structure of physiological signals to improve the quality of recovery.
The proposed data compressor and the recovery algorithm are validated on a dataset recorded on $12$ subjects during fast running.
Experiment results showed that an averaged $2.596$ beat per minute (BPM) estimation error was achieved
by jointly using compressed sensing with $50$\% compression ratio and a $2$-bit quantizer.
%The Pearson correlation between the heart rate estimates from the reconstructed datasets and the ground truth was $0.988$.
The results imply that we can effectively transmit $n$ bits instead of $n$ samples, which is a substantial improvement for
low-energy wireless telemonitoring.
\end{abstract}


\ifCLASSOPTIONpeerreview
\begin{center} \bfseries EDICS Category: SAS-MALN \end{center}
\fi
\IEEEpeerreviewmaketitle

%-------------------------------------------------------------------------

%---- 1. intro
\section{Introduction}
%-- 1.1 Background
%-- 1.2 Compression via CS, Analog Compression, Digital Compression
%-- 1.3 Recover Algorithms
%-- 1.4 Our contribution
%-- 1.5 Outline of the paper, denotations, etc.
In wireless health monitoring systems \cite{meier2013ehealth}, large amount and various types of physiological signals are collected
from on-body sensors, and then transmitted to nearby smart-phones or data central via wireless networks.
Energy consumption is a critical issue in these systems.
Compressed sensing (CS) \cite{Candes2008a} is a promising technique for such systems
for low-energy data acquisition, compression, and wireless transmission \cite{Zhang_TBME2012b, Craven2014}.

In the framework of CS, a signal $\ve{x}\in\mathbb{R}^N$ is compressed by a simple matrix-vector multiplication,
\begin{equation}\label{eq:compress_via_cs}
    \ve{y} = \bm{\Phi}\ve{x} + \ve{n}.
\end{equation}
where $\bm{\Phi}\in\mathbb{R}^{M\times N}$ is called the sensing matrix,
$\ve{y}\in\mathbb{R}^{M}$ is the compressed measurements and $\ve{n}$ is the measurement noise.
Usually $\bm{\Phi}$ is underdetermined, i.e., $M < N$, and the ratio $(N-M)/N$ is called the compression ratio of CS.

\begin{figure}[ht!]
\centering
\includegraphics[width=3in]{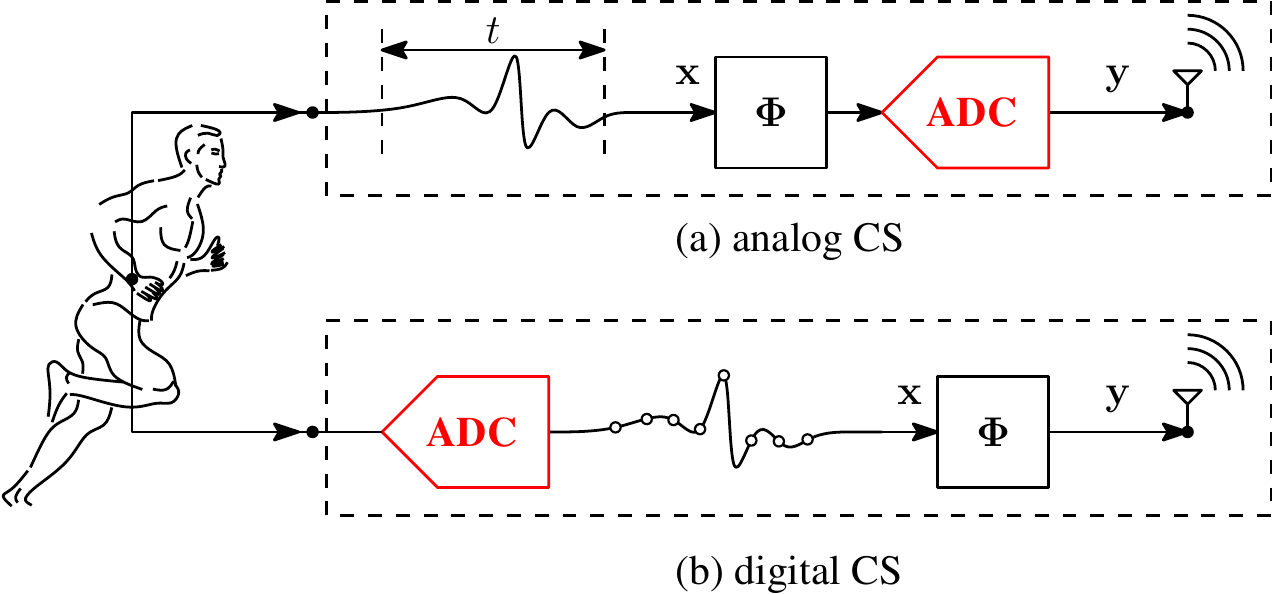}
\caption{Compression via Compressed Sensing (CS).
    The matrix-vector multiplication in \eqref{eq:compress_via_cs}
    can be either implemented using analog circuit in (a) or digitally in (b).
    In both cases, an ADC is required to convert continuous signals to quantized numbers.}
\label{fig:diagram}
\end{figure}

Compression via CS can be implemented in two ways as shown in Fig. \ref{fig:diagram}.
One is analog CS\cite{chen2012design, chen2013energy, haboba2012pragmatic, Wang2015},
where the matrix-vector multiplication $\bm{\Phi}\ve{x}$ is implemented in analog domain and the compressed measurements are
quantized via an Analog-to-Digital Converter (ADC).
Analog CS is also called Analog-to-Information Converter (AIC)\cite{haboba2012pragmatic, Wang2015}.
Usually, an AIC is implemented in a dedicate ASIC chip to achieve low-energy.
The other is digital CS, where signals are firstly quantized into digitalized samples and $\bm{\Phi}\ve{x}$ is calculated
in MCU\cite{mamaghanian2011compressed} or FPGA\cite{liu2013energy}.
The advantage of digital CS is that it can utilize exiting ADCs and can be implemented
as an efficient, low-energy data compressor\cite{liu2013energy}.
In both cases, the sensing matrix $\bm{\Phi}$ needs not to be transmitted and is stored in both the transmitter and receiver.

%A CS algorithm \cite{Candes2008a} can recover a signal from compressed measurements by utilizing the sparseness of the signal.
%A signal $\ve{x}\in\mathbb{R}^N$ is called $K$ sparse in basis $\bm{\Psi}\in\mathbb{R}^{N\times N}$,
%if we can drop all but the $K$ largest coefficients in $\ve{x}=\bm{\Psi}\bm{\theta}$, while still obtain an approximation to the original
%signal $\norm{\ve{x}^* - \ve{x}} = \norm{\bm{\Psi}\bm{\theta}^* - \bm{\Psi}\bm{\theta}}<\epsilon$.
%Eq \eqref{eq:compress_via_cs} is called a compressor where the number of compressed measurements $M$ is proportional\cite{Pfander2013}
%to the sparseness $K$.

At a receiver, the original signal $\mathbf{x}$ is recovered from the compressed measurements $\ve{y}$ via
\begin{equation}\label{eq:bp}
    \min \norm{\ve{x}}_0 \quad \text{subject to} \quad \| \ve{y} - \bm{\Phi}\ve{x} \|_2 \leq \varepsilon
\end{equation}
where $\norm{\ve{x}}_0$ is the $\ell_0$ norm of $\norm{\ve{x}}$, and $\varepsilon$ is the tolerance of noise or modeling errors.
Calculating the solution is very hard. Generally, one seeks the solution of a relaxed convex optimization problem \cite{becker2011nesta},
in which $\norm{\ve{x}}_0$ is  replaced with $\norm{\ve{x}}_1$ or other terms encouraging sparse solutions.
The quality of signal recovery can be improved
by exploiting other structure information in signals, such as wavelet-tree structure\cite{Baraniuk2010},
piecewise smooth\cite{Pant2014} or block sparse structure\cite{Zhang2012a, Zhang_TBME2012b}.

CS has been successfully used in low-energy telemonitoring of physiological signals such as EEG\cite{Zhang_TBME2012a} or fetal ECG\cite{Zhang_TBME2012b}.
However, these works\cite{Zhang_TBME2012a, Zhang_TBME2012b} assumed that the compressed measurements $\ve{y}$ was real-valued.
But in practice, the compressed measurements $\ve{y}$ must be quantized before transmission, i.e.,
\begin{equation}
    \ve{y} = \mathcal{Q}(\bm{\Phi}\ve{x})  \label{label:quantization}
\end{equation}
where $\mathcal{Q}(\cdot)$ is the quantization operator that maps a real-valued signal to finite quantization levels\cite{Gray1998}.
This process unavoidably introduces errors, called quantization errors.
Some CS algorithms \cite{Zymnis2010, Jacques2013, Yang2013b} were developed to recover signals from the quantized measurements.
This recovery procedure is often called de-quantization\cite{jacques2011dequantizing} and
a CS algorithm is also called a decoder\cite{Cambareri2013}.
Haboba \textit{et al.} studied the quantization errors and its effect on signal recovery using synthetic signals with fixed sparsity\cite{haboba2012pragmatic}.
However, the signal model used in \cite{haboba2012pragmatic} is not practical for telemonitoring applications, as most physiological signals are only approximately sparse.
Wang \textit{et al.} analyzed the performance-to-energy
trade-offs introduced by quantized CS\cite{Wang2015} in EEG telemonitoring
and provided a brute force method searching the optimum configuration of the quantizer.
But the CS algorithm used in \cite{Wang2015} ignored the model of quantization errors during signal recovery,
which may yield degraded performance.

Although quantization introduces errors, suitably using quantization can largely reduce the wireless transmission bit-budget\cite{laska2011democracy}.
%Unfortunately, this issue is rarely discussed in the literature on compressed sensing of physiological signals.
In this work, we address this issue and use quantized CS for low-energy telemonitoring.
First, physiological signals are compressed according to (\ref{eq:compress_via_cs}),
and then quantized according to (\ref{label:quantization}) with $2$ bits per measurement.
This compression scheme greatly reduces the transmission bit-budget, which benefits to low-energy telemonitoring.

On the de-compression and de-quantization stage, we propose a Bayesian de-quantization algorithm, denoted by BDQ. It exploits correlation structure within physiological signals and also takes into account the quantization errors. Note that this algorithm does not exploit sparsity to recover signals, as done by most other compressed sensing algorithms. Instead, it exploits correlation of physiological signals. Our motivation is that during wireless health monitoring many raw physiological signals are usually less sparse, namely these noisy signals are not sparse in the time domain and also not sparse in other transform domains \cite{Zhang_Asilomar}. In this case exploiting sparsity may not be very effective. In contrast, exploiting correlation may be a better direction, as shown in \cite{Zhang_TBME2012b}. However, the work in \cite{Zhang_TBME2012b} did not consider the quantization errors, and thus has inferior performance to our proposed BDQ algorithm, as shown in experiments later.

We study the application of Photoplethysmography (PPG) and accelerometer telemonitoring for fitness training,
in which the heart rate must be accurately estimated during intensive physical exercises.
We exploit the optimum compression by jointly tunning the compression ratio (CR) of CS and the quantization bit-depth.
The experiment results show that an averaged $2.596$ BPM absolute heart rate estimation error is achieved with
$\mathrm{CR}=0.50$ and a $2$ bits quantizer.
The Pearson correlation between the estimated heart rate from the recovered datasets and the ground-truth is $0.9899$.
These results imply that we can effectively compress raw segments of PPG and accelerometer data from $N$ samples into $N$ bits for
low-energy telemonitoring.

The rest of the paper is organized as follows. Section II introduces the quantization model of compressed sensing.
The Bayesian De-Quantize algorithm are presented and discussed in Section III.
Section IV describes the experimental set up and numerical results.
Discussions are given in Section V and Section VI concludes the paper.

%Throughout the paper, \textbf{Bold} letters are reserved for vector $\ve{x}$ and matrix $\ve{X}$.
%$\mathrm{Tr}(\ve{A})$ denotes the trace of a matrix $\ve{A}$.

%---- 2. two-stage data compression
\section{The Quantization Model of Compressed Sensing}

\begin{figure}[!ht]
    \centering
    \includegraphics[width=2.5in]{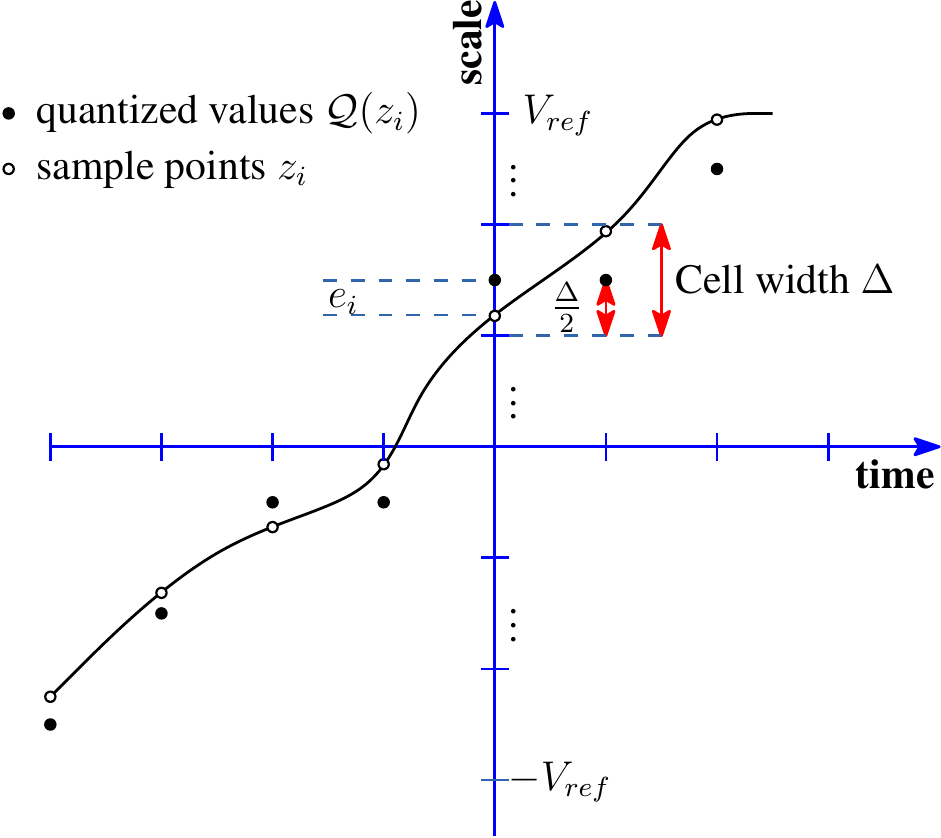}
    \caption{Time sampling and scale quantization. The quantization error is denoted by $e_i$ and the cell width is $\Delta$. The mid-point value within a cell is taken as the quantized value for a sample point falling in that cell.}
    \label{fig:adc}
\end{figure}

The function of an ADC is \textit{time} sampling and \textit{scale} quantization, which is illustrated in Fig. \ref{fig:adc}.
Let $B$ denotes the number of bits per measurement, which is also called the bit-depth of a quantizer.
Represented by $B$ bits, the scales between the positive and negative reference voltage $[-V_{ref}, V_{ref}]$ are
divided into $L=2^B$ quantization levels.
The cell width\cite{Gray1998} $\Delta$ of a uniform quantizer is
\begin{equation}\label{eq:adc_quantizer}
    \Delta = \frac{2V_{ref}}{2^B}.
\end{equation}
Signals larger than $V_{ref}$ or smaller than $-V_{ref}$ are saturated.
Those saturation signals are quantized with the same level as signals lie in
$[V_{ref}-\Delta, V_{ref}]$ or $[-V_{ref}+\Delta, -V_{ref}]$.
%Eq \eqref{eq:adc_quantizer} is a generalized model for an ADC which allows saturation.

For a sample $v_i\in\mathbb{R}$ falls in a cell, the mid-point value in that cell is used as the quantization value $\mathcal{Q}(v_i)$.
The quantization error $e_i$ is defined as
\begin{equation}
    e_i = \mathcal{Q}(v_i) - v_i,
\end{equation}
$e_i$ distributes uniformly between $[-\Delta/2, \Delta/2]$. Let $\mathcal{D}_e=[-\Delta/2, \Delta/2]$ denotes the
domain of quantization errors, then
\begin{equation}
    p(e_i) \sim \mathcal{U}(\mathcal{D}_e).
\end{equation}

It should be noted that $\mathcal{D}_e$ is unbounded when a signal saturates\cite{Yang2013b}.
Furthermore, the variance of the quantization error, denoted by $\sigma_e$, is,
\begin{equation}\label{eq:quantize_sigma_e}
    \sigma_e = \frac{\Delta^2}{12}.
\end{equation}

%----
\subsection{Analog CS}
For analog CS, the matrix-vector multiplication $\bm{\Phi}\ve{x}$ is implemented in analog domain and the
compressed measurements are quantized via an ADC before transmission,
\begin{equation}\label{eq:analog_cs_1}
    \ve{y} = \bm{\Phi}\ve{x} + \ve{n}, \quad \ve{y}_q = \mathcal{Q}(\ve{y})
\end{equation}
where the subscript $q$ denote the quantized signals. Let $\ve{e} = \ve{y}_q - \ve{y}$, we have,
\begin{equation}\label{eq:analog_cs_2}
    \ve{y}_q = \bm{\Phi}\ve{x} + \ve{n} + \ve{e}.
\end{equation}
The transmission bit-budget of analog CS is $MB$ bits, which can be controlled by the
number of compressed measurements $M$ as well as the quantization bit-depth $B$ of the ADC.
%To avoid saturation, an Automated Gain Control (AGC) circuit is usually used before ADC.
%----
\subsection{Digital CS}
For digital CS, analog signals are firstly quantized by an ADC via Nyquist sampling,
then the digitalized signals are compressed in MCU\cite{mamaghanian2011compressed} or FPGA\cite{liu2013energy} via a matrix-vector multiplication,
\begin{equation}\label{eq:digital_cs_1}
    \ve{y} = \bm{\Phi}\ve{x}_q + \ve{n}, \quad \ve{x}_q = \mathcal{Q}(\ve{x})
\end{equation}

In practice, larger bit-depth $B_i$ is required for the ADC to reduce the distortions during signal acquisition (i.e., $B_i=12$ was used in \cite{liu2013energy}).
For larger $B_i$, the variance of the quantization error $\ve{e}$ is much smaller and
\eqref{eq:digital_cs_1} is simplified as,
\begin{equation}\label{eq:digital_cs_3}
    \ve{y} = \bm{\Phi}\ve{x}_q + \ve{n}, \quad \ve{x}_q \approx \ve{x}
\end{equation}
%----
\subsection{The two-stage data compressor}
In \eqref{eq:digital_cs_1}, both $\ve{y}$ and $\ve{x}_q$ are represented in fixed-point arithmetic with $B_i$ bits\cite{liu2013energy}.
We may further reduce transmission bit-budget by simply rounding each sample of $\ve{y}$ to $B$ bits, denoted by $\ve{y}_q = \mathcal{Q}_r(\ve{y})$,
\begin{figure}[!ht]
    \centering
    \includegraphics[width=3.3in]{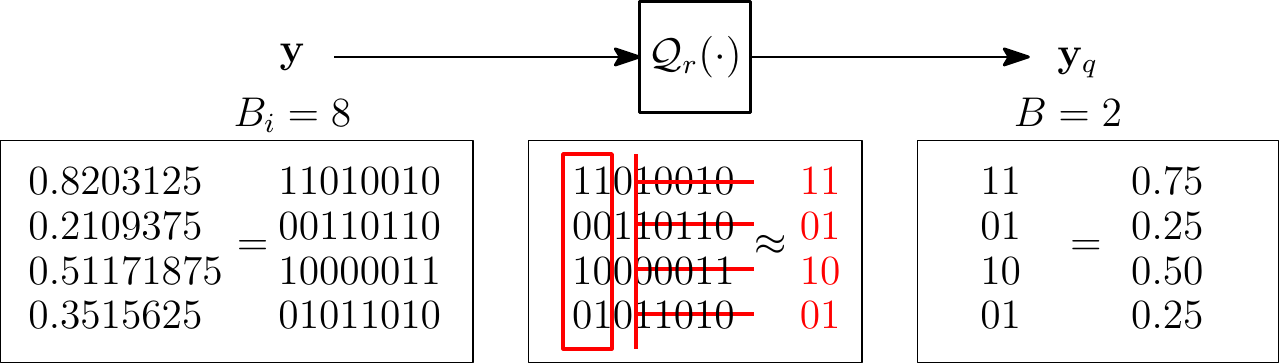}
    \caption{The operator $\mathcal{Q}_r(\cdot)$ can be efficiently implemented in digital CS.
In this example, the compressed measurements $\ve{y}$ are represented using unsigned fix-point arithmetic
($B_i=8$ and the fraction length is also $8$ bits). The rounding outputs $\ve{y}_q$ have only $2$ bits per sample.}
    \label{fig:rounding_op}
\end{figure}
where $\mathcal{Q}_r(\cdot)$ is called a rounding operator and its function is illustrated in Fig. \ref{fig:rounding_op}.
$\mathcal{Q}_r$ can be regarded as an economy quantizer in fixed-point arithmetic.

We denote by $\ve{e}_r=\ve{y}_q - \ve{y}$ the rounding error, which assumed to be uniform distributed,
\begin{equation}
    p(e_r) \sim \mathcal{U}(\mathcal{D}_{e_r})
\end{equation}
where $\mathcal{D}_{e_r}=[-\Delta_r/2, \Delta_r/2]$ and the cell width $\Delta_r = \frac{2V_{ref}}{2^{B}}$.
After rounding, \eqref{eq:digital_cs_3} can be reformulated as,
\begin{equation}\label{eq:digital_cs_2}
    \ve{y}_q = \bm{\Phi}\ve{x}_q + \ve{n} + \ve{e}_r, \quad \ve{x}_q\approx\ve{x}
\end{equation}

The bit-compression ratio $\mathrm{CR}_b$, i.e., the ratio between the reduced bit-budget after compression
divided by the total input bit-budget, is defined as
\begin{equation}\label{eq:bit_cr}
    \mathrm{CR}_b = \frac{NB_i - MB}{NB_i} = 1 - (1-\mathrm{CR})\cdot\frac{B}{B_i}
\end{equation}
where $\mathrm{CR}=(N-M)/N$ is the compression ratio in CS-based telemonitoring\cite{Zhang_TBME2012b, liu2013energy}.

\begin{figure}[!ht]
    \centering
    \includegraphics[width=3.4in]{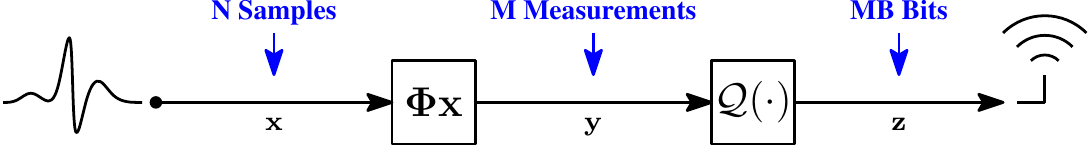}
    \caption{A unified framework for low-energy wireless telemonitoring using quantized compressed sensing.}
    \label{fig:cover_fig}
\end{figure}

In our paper thereafter, we do not distinguish between quantization $\mathcal{Q}(\cdot)$ in analog CS \eqref{eq:analog_cs_2} and
rounding $\mathcal{Q}_r(\cdot)$ in digital CS \eqref{eq:digital_cs_2}.
Instead, we model them in a unified framework in Fig. \ref{fig:cover_fig},
which is a two-stage data compressor that can be formulated as,
\begin{equation}\label{eq:qbsbl_model}
    \ve{z} = \bm{\Phi}\ve{x} + \ve{e} + \ve{n}
\end{equation}
where $\ve{z}$ is the quantized measurements represented with $B$ bits per sample, $\bm{\Phi}\in\mathbb{R}^{M\times N}$ is the sensing matrix,
$\ve{n}$ and $\ve{e}$ are measurement noise and quantization noise respectively.
In this framework, we transmit only $MB$ bits instead of $N$ samples for wireless telemonitoring.

%---- 3. Bayesian De-Quantize
\section{The Bayesian De-Quantize Algorithm}

\subsection{Bayesian Hierarchical Model}

%----
\subsubsection{Noise Model}
The measurement noise $\ve{n}$ is usually assumed Gaussian with variance $\lambda$, i.e., $\ve{n}\sim\mathcal{N}(\ve{0}, \lambda\ve{I})$.
From \eqref{eq:qbsbl_model} we have
\begin{equation}
    p(\ve{z} | \ve{x}, \ve{e}; \lambda) = \mathcal{N}(\bm{\Phi}\ve{x} + \ve{e}, \lambda\ve{I})
\end{equation}

The quantization error $\ve{e}$ is uniform distributed,
\begin{equation}
    p(\ve{e}) = \mathcal{U}(\mathcal{D}_e)
\end{equation}
where $\mathcal{D}_e = [-\Delta/2, \Delta/2]$ and $\Delta$ is the cell width of a quantizer.
In practice, the cell width $\Delta$ is known a prior given $B$ and the reference voltage $V_{ref}$.

\textbf{Remark 1:} The quantization error models rounding errors between the analog input and the digitalized output.
It is non-linear especially for low-resolution ADCs and also depends on the amplitude and frequency of a signal.
Exploiting the dependencies between signals and quantization errors may improve the quality of recovery,
however it is difficult to do so\cite{Yang2013b}.
To simplify our model, we assume an uniform distributions for $\ve{e}$
and do not consider the dependency between the quantization error and the analog input.

\textbf{Remark 2:} We studied only multi-bit ($B\geq2$) quantized CS and do not consider the extreme case of 1-bit compressed sensing.
1-bit CS may have better recovery performances than multi-bit CS under the same compressed bit-budget $MB$\cite{Jacques2013}.
However, 1-bit CS loses scale information of signals, which challenges its use in low-energy telemonitoring applications.
We refer the reader to the literature on 1-bit compressed sensing\cite{Jacques2013} for more details.

%----
\subsubsection{Signal Model}
We assume a correlated structure within the signal\cite{Zhang2012a} where we model the prior of signal $\ve{x}$ as,
\begin{equation}\label{eq:qbsbl_signal_model}
    p(\ve{x} | \gamma, \ve{P}) = \mathcal{N}(\ve{0}, \gamma\ve{P})
\end{equation}
where $\gamma$ is a non-negative parameter controlling the variance of the signal $\ve{x}$,
$\ve{P}$ is a symmetric positive semi-definite matrix modeling the correlation structure of signal $\ve{x}$.
The diagonal entries of $\ve{P}$ are normalized to $1$s during iterative learning.

%----
\subsection{The Bayesian De-Quantization Algorithm}

We estimate $\{\ve{x}, \ve{e}, \lambda, \gamma, \ve{P}\}$ using their joint MAP estimator,
\begin{align*}
    \{\hat{\ve{x}}, \hat{\ve{e}}, & \hat{\lambda}, \hat{\gamma}, \hat{\ve{P}}\}
    = \arg\max \log p(\ve{x}, \ve{e}, \lambda, \gamma, \ve{P} | \ve{z}) \\
    &= \arg\max \log p(\ve{e} | \ve{z}, \ve{x}, \lambda, \gamma, \ve{P}) \cdot p(\ve{x}, \lambda, \gamma, \ve{P} | \ve{z})
\end{align*}

\begin{figure}[!ht]
    \centering
    \includegraphics[width=3.4in]{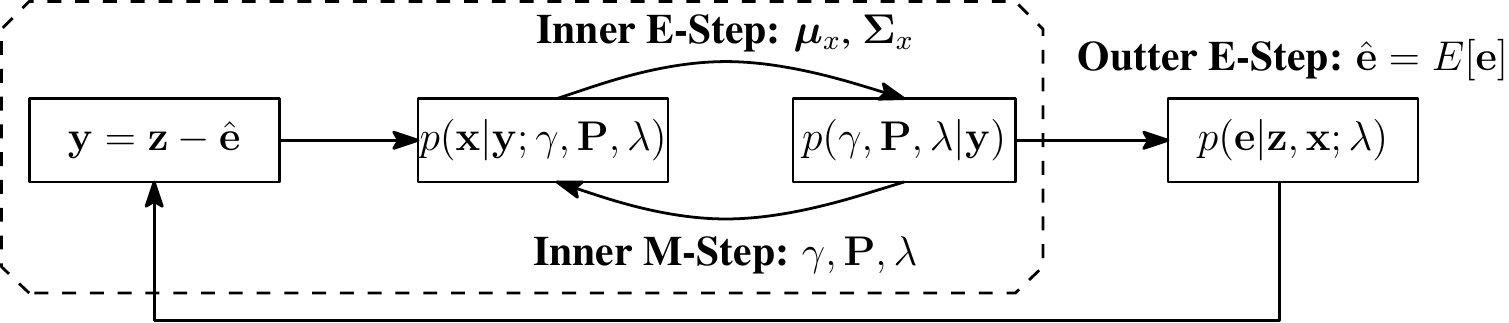}
    \caption{Bayesian De-Quantization Algorithm using Nested EM approach.}
    \label{fig:nestem}
\end{figure}

A nested Expectation Maximization (nest-EM) approach\cite{VanDyk2000, Prasad2014} is adopted.
The nest-EM is an iterative monotonically convergent method, it consists of an inner and outer EM loop,
which is briefly sketched in Fig. \ref{fig:nestem}.

We initialized the nest-EM procedure with $\ve{e}_0=\ve{0}$ and let $\ve{y} = \ve{z} - \ve{e}_0$, then iterative over,
\begin{itemize}
    \item \textbf{Inner E-Step}, estimate $\bm{\mu}_x$ and $\bm{\Sigma}_x$ from posterior probability $p(\ve{x}|\ve{y}; \gamma, \ve{P}, \lambda)$,
    \item \textbf{Inner M-Step}, update $\gamma$, $\ve{P}$ and $\lambda$ via maximize the likelihood $p(\gamma, \ve{P}, \lambda | \ve{y})$,
    \item \textbf{Outter E-Step}, estimate the first moment of quantization error $\ve{\hat{e}}$ from $p(\ve{e}|\ve{z}, \ve{x}; \lambda)$,
        and update $\ve{y} = \ve{z} - \ve{\hat{e}}$.
\end{itemize}

%----
\subsubsection{Inner-E step}
The posterior $p(\ve{x}|\ve{y}; \lambda, \gamma, \ve{P})$ can be expressed in an analytical form\cite{Tipping2001},
\begin{align*}
    & p(\ve{x}|\ve{y}; \lambda, \gamma, \ve{P}) = \frac{p(\ve{y}|\ve{x}; \lambda)p(\ve{x}|\gamma, \ve{P})}{p(\ve{y}|\lambda, \gamma, \ve{P})} \\
    & = (2\pi)^{-(N+1)/2}\abs{\bm{\Sigma}}^{-1/2}\exp\left\{ - \frac{1}{2}(\ve{x} - \bm{\mu})^T\bm{\Sigma}^{-1}(\ve{x} - \bm{\mu}) \right\},
\end{align*}
where $\bm{\mu}$ and $\bm{\Sigma}$ are respectively,
\begin{align}
    \bm{\mu} &= \gamma\ve{P}\bm{\Phi}^T(\lambda\ve{I} + \gamma\bm{\Phi}\ve{P}\bm{\Phi}^T)^{-1}\ve{y} \label{eq:qbsbl_mu} \\
    \bm{\Sigma} &= (\frac{1}{\gamma}\ve{P}^{-1} + \frac{1}{\lambda}\bm{\Phi}^T\bm{\Phi})^{-1} \label{eq:qbsbl_sigma}
\end{align}

%----
\subsubsection{Inner-M step}
The parameters $\lambda$, $\gamma$ and $\ve{P}$ are estimated by a Type II maximum likelihood procedure\cite{Tipping2001},
\begin{align}
& \mathcal{L}(\lambda, \gamma, \ve{P}) = - \log p(\ve{y}|\lambda, \gamma, \ve{P}) \\
& = \log\abs{\lambda\ve{I} + \gamma\bm{\Phi}\ve{P}\bm{\Phi}^T} + \ve{y}^T\left(\lambda\ve{I} + \gamma\bm{\Phi}\ve{P}\bm{\Phi}^T\right)^{-1}\ve{y}
\end{align}
Optimize over $\mathcal{L}(\lambda, \gamma, \ve{P})$, we have update rules for $\lambda$, $\gamma$ and $\ve{P}$ respectively,
\begin{align}
    \lambda &= \frac{\norm{\ve{y} - \bm{\Phi}\bm{\mu}}_2^2 + \mathrm{Tr}(\bm{\Sigma}\bm{\Phi}^T\bm{\Phi})}{M} \label{eq:qbsbl_4} \\
    \ve{P} &= \frac{\bm{\Sigma} + \bm{\mu}\bm{\mu}^T}{\gamma} \label{eq:qbsbl_p} \\
    \gamma &= \frac{1}{N} \mathrm{Tr}\left[ \ve{P}^{-1} (\bm{\Sigma} + \bm{\mu}\bm{\mu}^T) \right] \label{eq:qbsbl_gamma0}
\end{align}

\textbf{Remark 3:}
Due to the coupling of $\ve{e}$ and $\ve{n}$ in \eqref{eq:qbsbl_model}, the estimate of the noise variance $\lambda$ is inaccurate due
to an identifiability issue as stated in \cite{Zhang2011}. Therefore, $\lambda$ is often treated as a regularize parameter
\cite{liu2013energy, Yang2013b}. We set $\lambda=0.001$ as the default parameter used in the experiment.

%----
\subsubsection{Regularization on $\ve{P}$}
Regularization on $\ve{P}$ is required due to limited data\cite{Zhang2012a}. In \cite{Zhang2012a}, the author
provided an empirical method on the regularization of $\ve{P}$ using a symmetric Toeplitz matrix,
\begin{equation}\label{eq:ar1_zz_1}
    \ve{P}_{ij} = r^{\abs{i-j}}, \qquad i,j = 1,\cdots,N, \quad \abs{r}<1
\end{equation}
where $r$ is the correlation coefficient empirically calculated from the ratio between the mean of sub-diagonal of $\ve{P}$ and the
mean of main diagonal of $\ve{P}$. Such regularization is equivalent to modeling the correlation structure
as a first-order Auto-Regressive (AR) process\cite{Zhang2011}.

An AR(1) matrix has simple tri-diagonal inverse,
\begin{equation}\label{eq:inv_ar1}
\ve{P}^{-1} = \frac{1}{1-r^2}
\begin{pmatrix}
1  & -r     & 0      & \cdots & 0  \\
-r & 1+r^2  & -r     & \cdots & 0  \\
\vdots   & \ddots & \ddots & \ddots &  \vdots  \\
0  & \cdots & -r     & 1+r^2  & -r \\
0  & \cdots & 0      & -r     & 1  \\
\end{pmatrix}
\end{equation}
and it can be decomposed as
\begin{equation}
    \ve{P}^{-1} = \frac{1}{1-r^2}\ve{T}\left( \ve{D}^T\ve{D} \right)
\end{equation}
where $\ve{T}=\mathrm{diag}\{1,\cdots,1,1/(1+r^2)\}$ is a diagonal matrix
and $\ve{D}$ is a temporal smooth operator\cite{zhou2011multi} defined as
\begin{equation}\label{eq:ar1_zz_2}
\ve{D} =
\begin{pmatrix}
1 & -r     & 0      & \cdots & 0  \\
0 & 1      & -r     & \cdots & 0  \\
  & \ddots & \ddots & \ddots &    \\
0 & 0      & \cdots & 1      & -r \\
0 & 0      & \cdots & 0      & 1
\end{pmatrix},
\end{equation}
A similar regularization method was proposed in \cite{Babacan2012} which models the inverse covariance matrix as $\ve{D}^T\ve{D}$.

This type of regularization can exploit the correlated structures within
physiological signals\cite{Zhang_TBME2012b, Zhang_TBME2012a}.
However, the correlation coefficient $r$ in \eqref{eq:ar1_zz_1} or \eqref{eq:ar1_zz_2} can only be
empirically calculated\cite{Zhang2012a} or fixed\cite{Babacan2012} in existing algorithms.

We presented in this paper an AR(1) approximation to estimate the correlation matrix in \eqref{eq:qbsbl_p}
using Karhunen-Loeve Transform (KLT) \cite{Akansu2012}.
The correlation matrix $\ve{P}$ calculated by \eqref{eq:qbsbl_p} is a symmetric, positive semi-definite matrix. Let
\begin{equation}\label{eq:eig_decomp}
    \ve{U}\ve{D}\ve{U}^T = \mathrm{SVD}(\ve{P})
\end{equation}
denotes the Singular Value Decomposition (SVD) of $\ve{P}$,
where columns of $\ve{U}$ are the eigenvectors of $\ve{P}$ and
$\ve{D}$ is a diagonal matrix with $\ve{d}=\mathrm{diag}(\ve{D})$ being the descending ordered eigenvalues of $\ve{P}$.
It have been shown in \cite{Strang1999} that the basis vectors of a Discrete Cosine Transform (DCT) approach
to the eigenvectors of the inverse of an AR(1) matrix as the coefficient $r$ goes to $1$. More precisely, from \eqref{eq:inv_ar1}
we define
\begin{equation}\label{eq:eig_dct}
    \ve{K} \triangleq \ve{P}^{-1}(1-r^2),
\end{equation}
then the \textit{rows} of a DCT Type-2 matrix is the eigenvectors of $\ve{K}$
as $r\rightarrow1$. This property of DCT has made it a popular transform for decomposition of highly correlated signal sources\cite{Akansu2012}.

From \eqref{eq:qbsbl_p}, \eqref{eq:eig_decomp} and \eqref{eq:eig_dct},
we may regularize the correlation matrix $\ve{P}$ with an AR(1) matrix by simply substituting the eigenvalue matrix
$\ve{U}$ in \eqref{eq:eig_decomp} with a DCT Type-2 matrix,
\begin{equation}\label{eq:qbsbl_barp_1}
    \tilde{\ve{P}} = \mathrm{dctmtx}(N)^T\,\ve{D}\,\mathrm{dctmtx}(N)
\end{equation}
where $\mathrm{dctmtx}(N)$ generates a DCT Type-2 matrix of size $N$.
We then normalize the diagonal entries of $\tilde{\ve{P}}$ to $1$s by
\begin{equation}
    \ve{v} = \sqrt{\mathrm{diag}(\tilde{\ve{P}})}, \quad
    \bar{\ve{P}} = \mathrm{diag}^{-1}(\ve{v}) \tilde{\ve{P}} \mathrm{diag}^{-1}(\ve{v}) \label{eq:qbsbl_barp_2}
\end{equation}
where $\mathrm{diag}^{-1}(\ve{v})$ builds a diagonal matrix with the diagonal entries given by $\ve{v}$. In \eqref{eq:qbsbl_barp_2},
$\bar{\ve{P}}$ was the AR(1) approximation to the correlation matrix $\ve{P}$.

$\gamma$ was updated afterwards using the regularized $\bar{\ve{P}}$,
\begin{equation}
    \gamma = \frac{1}{N} \mathrm{Tr}\left[ \bar{\ve{P}}^{-1} (\bm{\Sigma} + \bm{\mu}\bm{\mu}^T) \right] \label{eq:qbsbl_gamma}
\end{equation}

\textbf{Remark 4:} Using KLT to exploit the correlation structure in signals has long been exist in literature.
%Often, the KLT was used implicitly under the names such as SVD, Principle Component Analysis (PCA), etc., as eigen decomposition methods.
In \cite{Akansu2012}, the author studied the use of Toeplitz matrix to approximate empirical correlation matrix.
To our best knowledge, our work was the first to use KLT in regularized least squares to solve underdetermined optimization problems.
The DCT approximation in \eqref{eq:qbsbl_barp_1} is also attractive for its computational efficiency
in calculating $\bm{\mu}$, $\bm{\Sigma}$ and $\gamma$ in \eqref{eq:qbsbl_mu}, \eqref{eq:qbsbl_sigma} and \eqref{eq:qbsbl_gamma}.
%where $\tilde{\ve{P}}^{-1} = \mathrm{dctmtx}(N)^T\,\ve{D}^{-1}\,\mathrm{dctmtx}(N)$ was two-dimensional inverse DCT applied on $\ve{D}^{-1}$.

%----
\subsubsection{Estimate The Quantization Error $\ve{e}$ (Outer-E Step)}
We calculate the expected value of $\ve{e}$ from
\begin{equation*}
    \ve{\hat{e}} = E[\ve{e}|\ve{e}\in\mathcal{D}_e] = \int_{-\Delta/2}^{\Delta/2} \ve{e}\;p(\ve{e}|\ve{z},\ve{x}; \lambda) \mathrm{d} \ve{e}
\end{equation*}
which is the expected value of a truncated Normal distribution.
$\ve{\hat{e}}$ can be obtained analytically\cite{Yang2013b},
\begin{equation}\label{eq:qbsbl_e}
    \ve{\hat{e}} = \bm{\mu}_e - \sqrt{\lambda} \cdot \frac{\mathrm{PDF}(\ve{l}_e) - \mathrm{PDF}(\ve{u}_e)}{\mathrm{CDF}(\ve{l}_e) - \mathrm{CDF}(\ve{u}_e)}
\end{equation}
where $\bm{\mu}_e = \ve{z} - \bm{\Phi}\bm{\mu}_x$, $\ve{l}_e = (-\Delta/2 - \bm{\mu}_e)/\sqrt{\lambda}$
and $\ve{u}_e = (\Delta/2 - \bm{\mu}_e) / \sqrt{\lambda}$.
$\mathrm{PDF}(\cdot)$ and $\mathrm{CDF}(\cdot)$ are the probability density function (PDF) and
cumulative density function (CDF) of a standard Normal distribution respectively.

\textbf{Remark 5:}
Besides the quantization cell width $\Delta$, we may also have the prior information of the reference voltage $V_{ref}$.
For the estimates of the measurements $\hat{\ve{z}} = \bm{\Phi}\bm{\mu}_x$, values larger than $V_{ref}$ might indicate that
saturation occurs. In this case, we may write \eqref{eq:qbsbl_e} as,
\begin{equation}
    \ve{\hat{e}}_{\mathcal{I}} = \bm{\mu}_{e\mathcal{I}} - \sqrt{\lambda} \cdot \frac{\mathrm{PDF}(\ve{l}_{e\mathcal{I}})}{\mathrm{CDF}(\ve{l}_{e\mathcal{I}}) - 1}
\end{equation}
where $\mathcal{I}$ denotes the set of index where $\hat{\ve{z}}>V_{ref}$.

\subsubsection{The proposed algorithm}
The resulting algorithm is summarized in Fig. \ref{algo:qbsbl},
named as the Bayesian De-Quantize algorithm (\textbf{BDQ}).

\begin{figure}[!ht]
\centering
\begin{algorithmic}[1]
    \algrule
\Procedure{BDQ}{$\ve{z}$,$\bm{\Phi}$}
\State Outputs: $\ve{x},\ve{e}$
\State Initialize: $\gamma=1$, $\ve{P}=\ve{I}$, $\lambda=0.001$
\While{not converged}
\State Estimate $\bm{\mu}$, $\bm{\Sigma}$ by \eqref{eq:qbsbl_mu}, \eqref{eq:qbsbl_sigma}.
\State Calculate $\ve{P}$ via \eqref{eq:qbsbl_p}
\State Regularize $\bar{\ve{P}}$ via \eqref{eq:qbsbl_barp_1}-\eqref{eq:qbsbl_barp_2}.
\State Calculate $\gamma$ via \eqref{eq:qbsbl_gamma}.
\State Calculate $\hat{\ve{e}}$ by \eqref{eq:qbsbl_e}
\State Update $\ve{y} = \ve{z} - \hat{\ve{e}}$
\EndWhile
\EndProcedure
    \algrule
\end{algorithmic}
\caption{The Bayesian De-Quantize (BDQ) algorithm.}
\label{algo:qbsbl}
\end{figure}

\textbf{Remark 6:}
The \textbf{BDQ} algorithm can be used to recover piecewise smooth signals from quantized, and possibly underdetermined measurements.
It shares some similarities with the Block Sparse Bayesian Learning (BSBL) framework\cite{Zhang2012a} in non-sparse mode
where only the correlations within signals are exploited.
However, in the experiment we found that the regularization method \eqref{eq:qbsbl_barp_1}-\eqref{eq:qbsbl_barp_2}
on $\ve{P}$ in \textbf{BDQ} was superior to the empirical methods in BSBL,
which yielded better recovery results on physiological signals.

%---- 4. Experiment
\section{Experiments and Results}

%----
\subsection{Datasets}
\begin{figure}[!ht]
    \centering
    \includegraphics[width=3.4in]{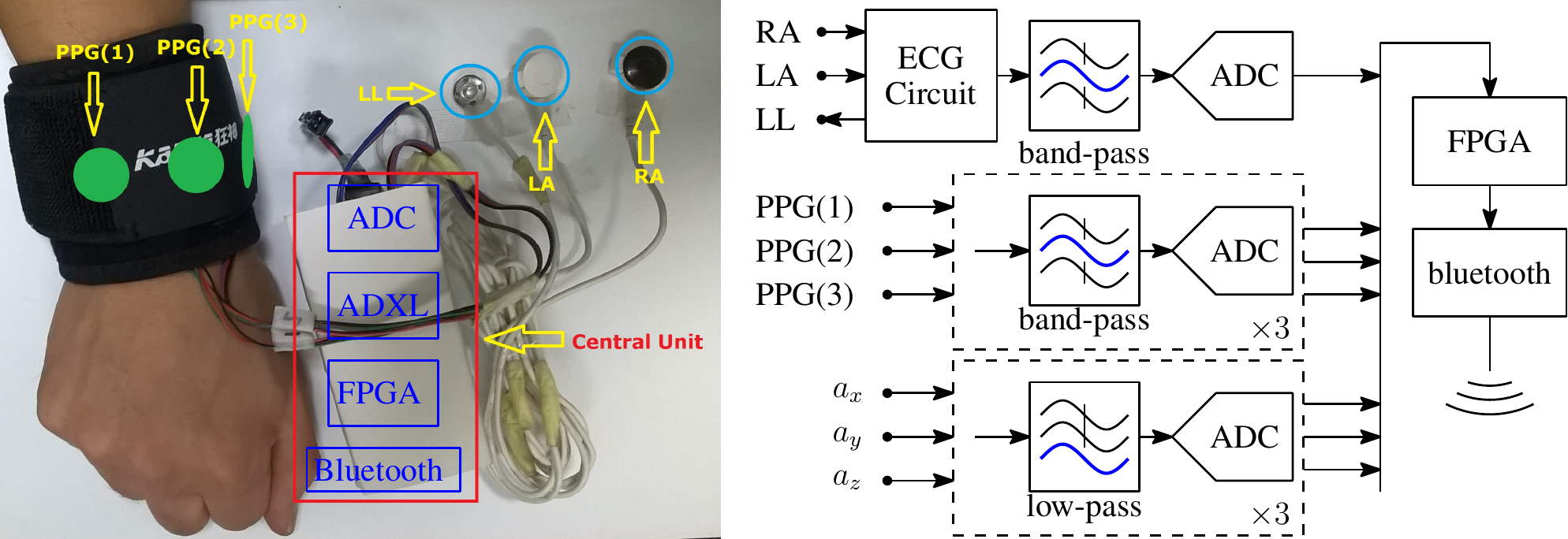}
    \caption{The hardware setup for data recording. ECG, PPG and accelerometer signals were simultaneously collected.
        Wet electrodes were used to obtain a three-lead ECG,
        where the leads LA and RA were placed at left and right chest respectively, and the LL lead was placed at the left lower abdomen.
        PPG signals were collected using reflective pulse oximeter sensors with green LED (wavelength: 515nm).
        Two PPG sensors were placed at the back of the wrist while another one was placed right on the pulse position.
        Three axial accelerometer data were collected using ADXL345 with $\pm4$g range.
        Data were transmitted to a laptop via bluetooth.}
    \label{fig:hardware}
\end{figure}

We simultaneously collected ECG, PPG and accelerometer signals from $12$ volunteers with age ranged from $18$ to $35$.
Fig. \ref{fig:hardware} shows the hardware set up for data recording.
%A reflective type PPG sensor was used. It illuminates green LED (wavelength = $515$nm) and detects the reflected light intensity
%using a photon sensor with peak sensitivity at 565nm.
For each subject, we recorded data for $5\sim6$ minutes. During data collection, a subject ran on a treadmill with speeds ranged
from $2$km/hour to $15$km/hour. The sensor band was conveniently worn on the wrist
and we intentionally introduced additional artifacts by asking all subjects to pull clothes, wipe sweat, swig arms during data recording.

All signals were filtered and then sampled at $125$Hz with $12$ bits precision per sample.
Table \ref{tab:filters} shows the specifications for all analog filters.
\begin{table}[!ht]
    \renewcommand{\arraystretch}{1.3}
    \centering
    \caption{Specifications for analog filters. $f_{c1}$ and $f_{c2}$ denote the
    lower and upper cutoff frequencies at $22$dB attenuation.}\label{tab:filters}
    \begin{tabular}{cccc}
        \toprule
        & ECG & PPG & Accelerometer \\
        \midrule
        Filter Type & Band-pass & Band-pass & Low-pass \\
        $f_{c1}$ & $0.1$Hz & $0.25$Hz & -- \\
        $f_{c2}$ & $100$Hz & $14.5$Hz & $100$Hz \\
        \bottomrule
    \end{tabular}
\end{table}

In the simulation framework, ECG signals were only used as reference signals to extract the ground truth heart rate.
One channel PPG (site PPG(1) placed at the back of the wrist) and there-axis accelerometer were the data actually used.
We downsampled (Nyquist sampling) PPG and accelerometer signals to $31.75$Hz in the experiments.
It should be noted that the accelerometer data contained aliasing when downsampled at $31.75$Hz.
However, such distortions were small as the rate of most activities during fitness training rarely exceed $10$Hz.

%----
\subsection{Experiments Setup}
\begin{figure}[!ht]
    \centering
    \includegraphics[width=3.3in]{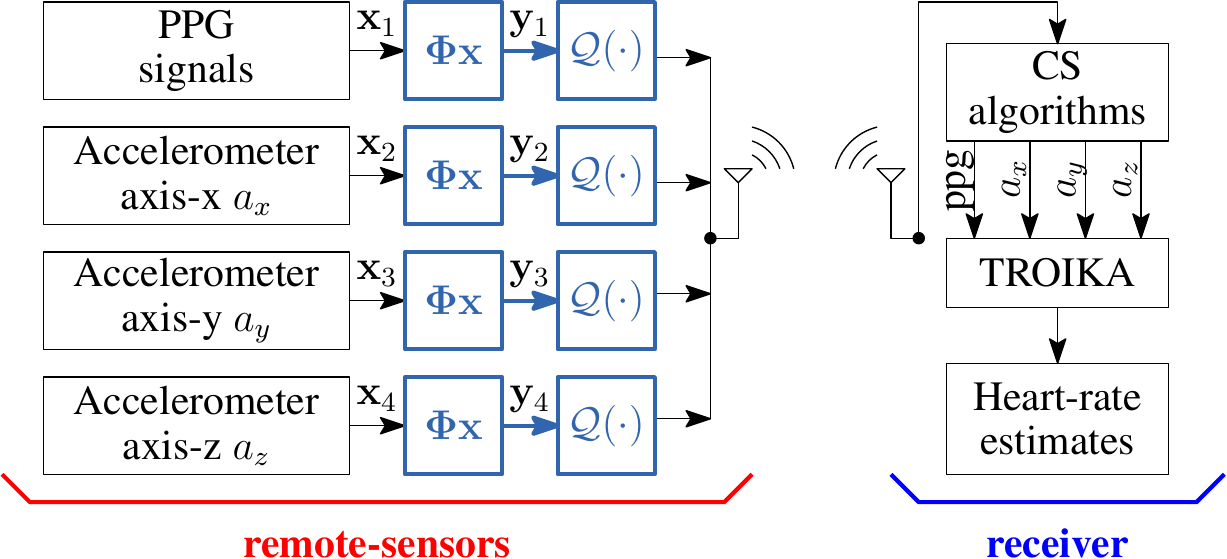}
    \caption{The quantized compressed sensing was applied on PPG and three axial accelerometer data to reduce the transmission bit-budget.
    At the receiver, data were recovered from quantized measurements, and the TROIKA framework
    was applied on the recovered data to estimate the heart rate.}
    \label{fig:troika_qcs}
\end{figure}

% illustrative
% bit-depth
% different quantizer [digital CS vs analog CS]
Fig. \ref{fig:troika_qcs} shows the experiment setup.
Quantized compressed sensing is applied on PPG and accelerometer signals.
Raw segments $\{\ve{x}_1, \ve{x}_2, \ve{x}_3, \ve{x}_4\}$ from each data channel are compressed
%\footnote{PPG signals usually have high dynamic range especially when the subject is under extensive exercise,
%often an automatic gain control circuitry is required before an ADC to avoid saturation (overflow) or underflow.
%In this experiment, we simplified the numerical simulation by signal normalization via $\ve{x} / \norm{\ve{x}}$
%before compression and quantization.}
by $\ve{y}=\bm{\Phi}\ve{x}$ simultaneously.
Sparse binary sensing matrices, whose entries consisted of only $0$s and $1$s, are used in the experiment.
Such matrix preserves low-power property when implemented in FPGA\cite{liu2013energy}.
We fix each column of $\bm{\Phi}$ consisting exactly $2$ non-zero entries and also make sure $\bm{\Phi}$ is full row-rank in each iteration.
The compressed measurements $\{\ve{y}_1, \ve{y}_2, \ve{y}_3, \ve{y}_4\}$ are further quantized by $\mathcal{Q}(\cdot)$ to
reduce the transmission bit-budget.

At the receiver, CS algorithms are used to recover signals from quantized measurements.
The heart rate is estimated by the TROIKA\cite{Zhang2015} framework using the recovered PPG and accelerometer datasets.

%Compression via compressed sensing usually operates in a batch mode
%\footnote{It can also be implemented in an online mode using $\ve{y}^{(i)} = \ve{y}^{(i-1)} + \bm{\phi}_ix_i$, where $\ve{y}^{(i)}$ is
%the compressed measurements after the sample $x_i$ has been acquired, $\bm{\phi}_i$ is the $i$th column of $\bm{\Phi}$. Readers may
%refer to \cite{liu2013energy, liu2013compression} for more details.}
%where signals are divided into equal-sized segments and each segment has $N$ samples.

The codes and data reproducing the results in the experiments are available at \url{https://github.com/liubenyuan/qsbl}.

%----
\subsection{Performance Measurement}
We use two performance metrics. One is the reconstruction SNR (RSNR),
\begin{equation*}
    \mathrm{RSNR}~(\text{dB}) = 10\log_{10} \frac{\norm{\ve{x}}_2^2}{\norm{\ve{\hat{x}} - \ve{x}}_2^2}
\end{equation*}
where $\hat{\ve{x}}$ denotes the recovered signal of $\ve{x}$. To assess the averaged performance over all segments,
we refer to the average RSNR (ARSNR), defined by
\begin{equation}
    \mathrm{ARSNR}~(\text{dB})=10\log_{10}\frac{1}{S}\sum_{i=1}^S \left( \frac{\norm{\ve{x}_i}_2^2}{\norm{\ve{\hat{x}_i} - \ve{x}_i}_2^2} \right)
\end{equation}
where $\ve{x}_i$ is the $i$th signal segment and $S$ is the total number of segments in a dataset.
The second metric is the Structural SIMilarity index (SSIM) \cite{Wang2009}. SSIM measures the similarity between recovered signals and
original signals, which is a better metric than RSNR\cite{Wang2009, Zhang_TBME2012a}.

The qualities of recovery are not only characterized by RSNR, but also by application specific requirements.
Therefore, we perform a task-driven approach where the heart rate estimates from the recovered PPG and accelerometer signals
are evaluated. The average absolute estimation error (Error1), defined in \cite{Zhang2015}, was,
\begin{equation}
    \mathrm{Error1} = \frac{1}{W}\sum_{i=1}^W \abs{\mathrm{BMP}_{\mathrm{est}}(i) - \mathrm{BPM}_{\mathrm{true}}(i)}
\end{equation}
where $W$ is the total number of heart rate estimates,
$\mathrm{BPM}_{\mathrm{est}}(i)$ is the estimated heart rate in the $i$th time window
\footnote{The TROIKA algorithm\cite{Zhang2015} operates in a sliding window manner. A time window of $T$ seconds
is sliding on the signals with incremental step $S$ seconds,
and the heart rate estimates are based on the samples collected within this time window. We
use default parameters ($T=8$s, $S=2$s) for TROIKA in our experiments.}
and $\mathrm{BPM}_{\mathrm{true}}(i)$ is the ground truth heart rate.
The standard deviation of heart rate estimates, denoted by $\mathrm{SD}_{\mathrm{BPM}}$, is
\begin{equation}
    \mathrm{SD}_{\mathrm{BPM}} = \sqrt{\frac{1}{W}\sum_{i=1}^W (\mathrm{BMP}_{\mathrm{est}}(i) - \mathrm{BPM}_{\mathrm{true}}(i))^2}
\end{equation}
Pearson correlation between the ground-truth and the heart rate estimates is also calculated for comparison.

%----
\subsection{The Recovery Algorithms for Quantized CS}
Besides the proposed algorithm, we use the following two typical CS algorithms: QVMP\cite{Yang2013b} and BSBL-BO\cite{Zhang_TBME2012b}.

\textbf{(1) QVMP\cite{Yang2013b}.} QVMP is a variational Bayesian De-Quantization algorithm proposed in \cite{Yang2013b}.
As shown in \cite{Yang2013b}, it has better performance than QIHT\cite{Jacques2013} and L1RML\cite{Zymnis2010}.
However, QVMP can not recover physiological signals directly in time domain. Instead, we applied QVMP in transformed domain, where
\[
    \ve{z} = (\bm{\Phi\Psi})\bm{\theta} + \ve{e} + \ve{n}
\]
$\bm{\Psi}$ is a sparse representation matrix and $\bm{\theta}$ is the sparse coefficients.
QVMP firstly recovered $\hat{\bm{\theta}}$, then $\hat{\ve{x}}$ via $\hat{\ve{x}}=\bm{\Psi}\hat{\bm{\theta}}$.

In the experiment, Discrete Cosine Transform (DCT) matrix was selected for $\bm{\Psi}$
and we set $\lambda=0.001/M$, $\mathrm{tol}=1e^{-6}$, $\mathrm{maxiter}=400$ for QVMP
as it achieved best recovery performance on the datasets.

\textbf{(2) BSBL-BO\cite{Zhang_TBME2012b, Zhang2012a}.}
BSBL-BO\cite{Zhang2012a} is the best performing CS algorithm in recovering physiological signals such as
fetal ECG\cite{Zhang_TBME2012b} or EEG\cite{Zhang_TBME2012a}.
%It is valuable to compare quantized CS algorithms with BSBL-BO that unaware of the quantization error.
In the experiment, we observed that BSBL-BO can also recover signals from quantized measurements,
by modeling quantization errors using a zero mean Normal distribution with a larger variance $\lambda_e$.

Throughout the experiment, BSBL-BO worked in non-sparse mode and directly recovered signals in time domain.
We selected $\mathrm{blkLen}=32$, $\mathrm{LearnLambda}=2$, $\mathrm{maxiter}=64$ and $\mathrm{LearnType}=1$
for BSBL-BO as this setting achieved best recovery performance.
The recoveries of BSBL-BO using real-valued measurement were also calculated, where the parameter $\mathrm{LearnLambda}=0$ was used.

% place here for better visualization
\begin{figure*}[!ht]
    \centering
    \includegraphics[width=0.99\textwidth]{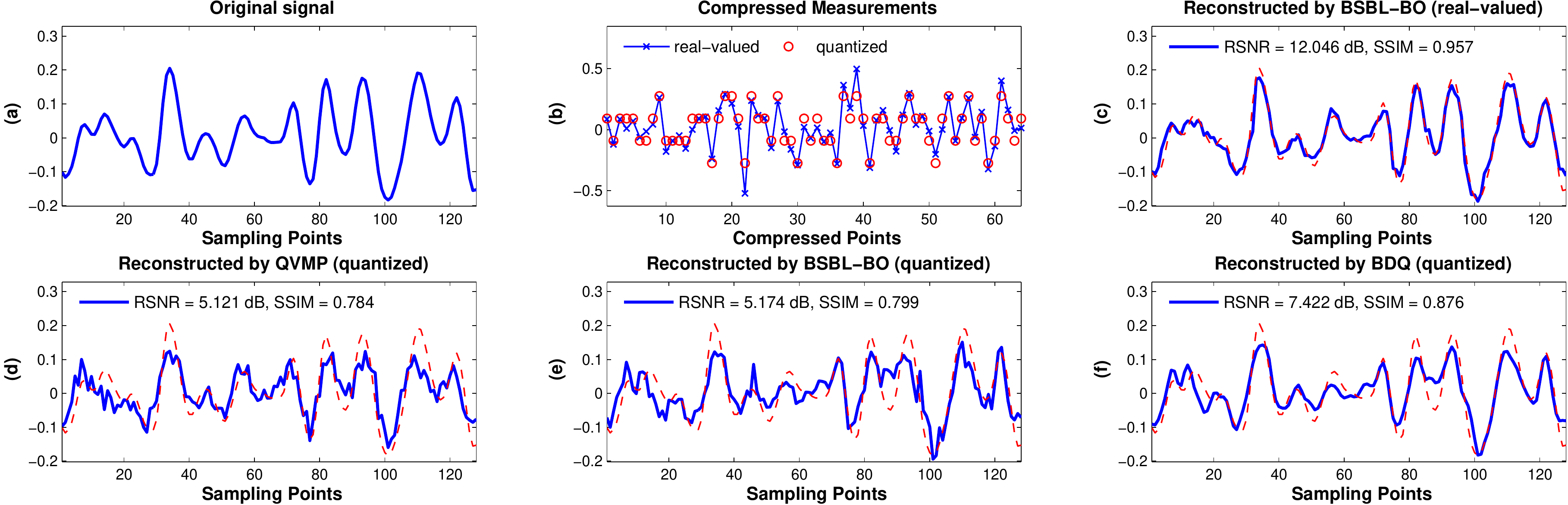}
    \caption{Performances of different CS algorithms on quantized measurements.
        (a) a raw PPG segment $\ve{x}$ was collected in $4$s ($N=128$).
        (b) real-valued compressed measurements ($M=64, \mathrm{CR}=0.50$) were generated via $\ve{y}=\bm{\Phi}\ve{x}$ and
        then quantized by $\mathcal{Q}(\cdot)$ with $2$ bits per measurement. The transmission bit-budget was only $MB=128$ bits.
        (c) on real-valued measurements, BSBL-BO recovered original signals with $\mathrm{RSNR}=12.046$ dB and $\mathrm{SSIM}=0.957$.
        (d) QVMP recovered signals in DCT domain with $\mathrm{RSNR}=5.121$ dB and $\mathrm{SSIM}=0.784$.
        (e)(f) BSBL-BO and BDQ recovered signals directly in time domain with $\mathrm{RSNR}=5.174$ dB, $\mathrm{SSIM}=0.799$ and
        $\mathrm{RSNR}=7.422$ dB, $\mathrm{SSIM}=0.876$ respectively.}
    \label{fig:figure01}
\end{figure*}

%---- regen 0124
\subsection{Results}

%-------- exp 1 --------
\subsubsection{An illustrative example}
\begin{figure}[!ht]
    \centering
    \subfloat[ARSNR]{\includegraphics[width=.49\linewidth]{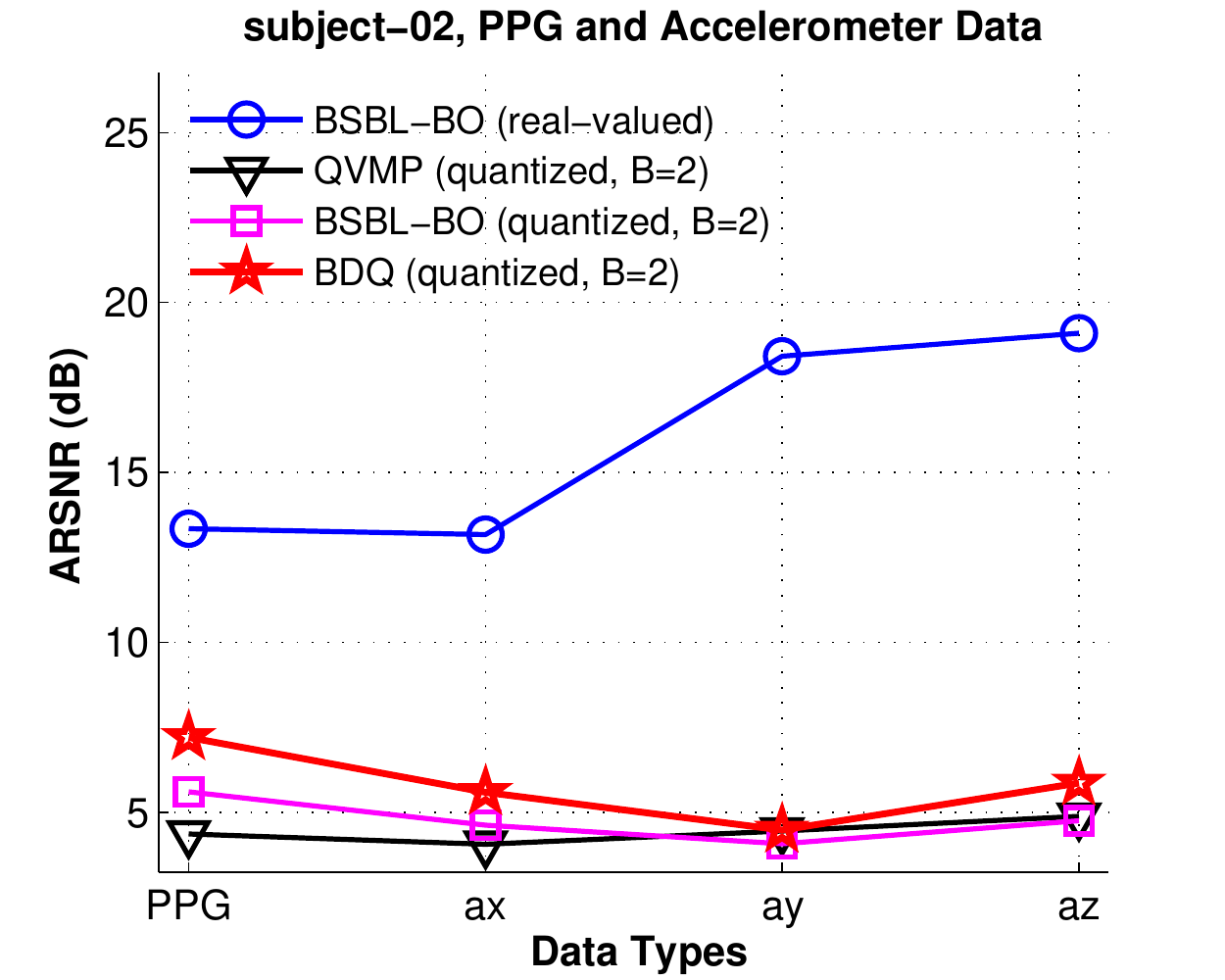}}
    \subfloat[SSIM]{\includegraphics[width=.49\linewidth]{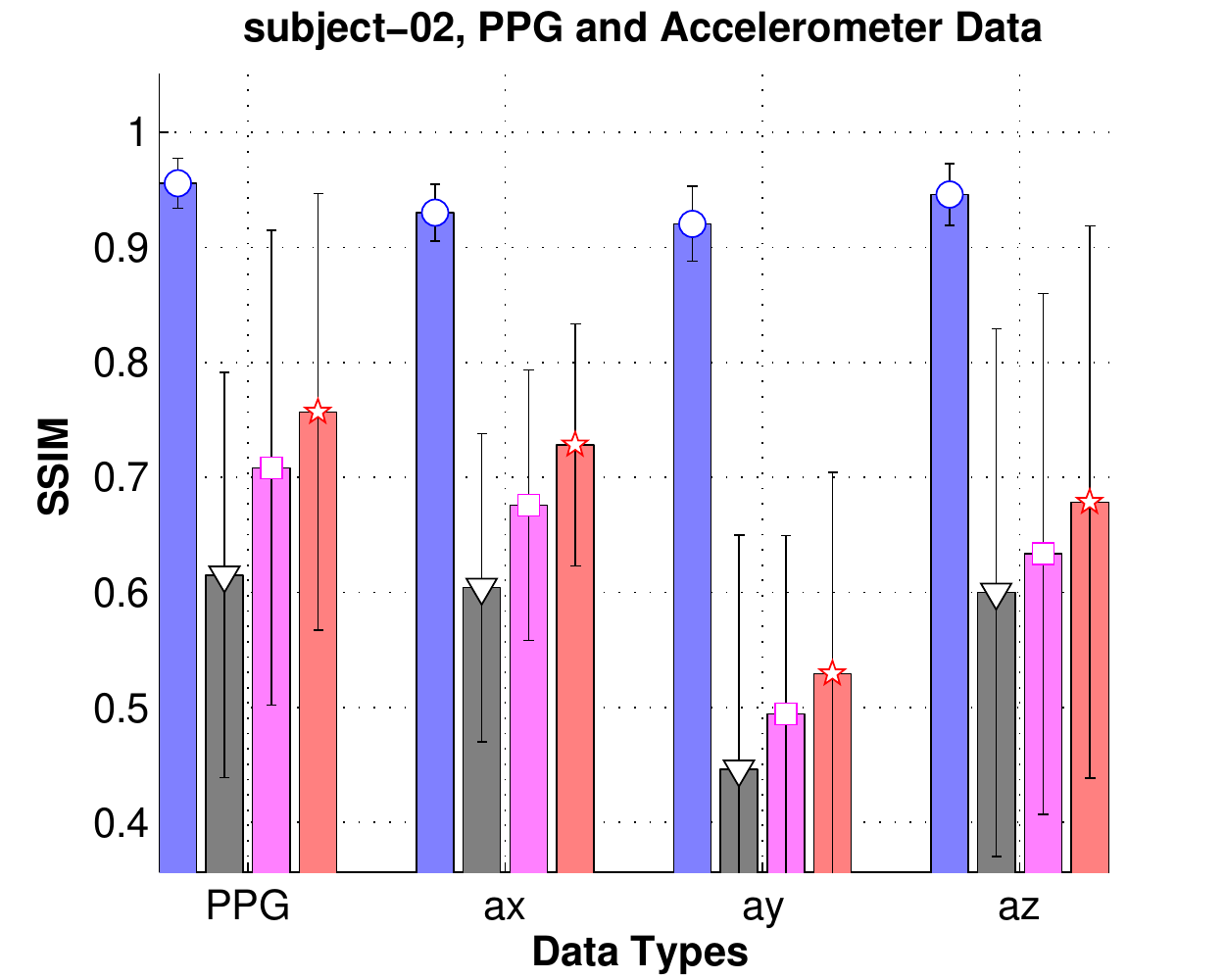}}
    \caption{The ARSNR and SSIM of signal recoveries w.r.t different types of signals. The parameters $N=128$, $M=64$, $B=2$ were used.
    The three-axial accelerometer data were denoted by ax (x-axis), ay (y-axis) and az (z-axis) respectively.}
    \label{fig:figure02}
\end{figure}

To better understand the quantized compressed sensing and the quality of signal recovery using different CS algorithms,
an illustrative example was given in Fig. \ref{fig:figure01}.
\begin{enumerate}
\item Fig. \ref{fig:figure01} (a). A raw PPG segment $\ve{x}$ of size $N=128$ was shown. In our experiments, signals were
    divided into fix-sized segments and each segment was normalized (i.e., $\ve{x}/\norm{\ve{x}}$) before compression.
\item Fig. \ref{fig:figure01} (b). A segment $\ve{x}$ was compressed via CS, $\ve{y} = \bm{\Phi}\ve{x}$, where $\bm{\Phi}$ was a sparse
    binary matrix with $M=64$ rows and the compression ratio $\mathrm{CR}=0.50$. The compressed measurements were further quantized
    by $\mathcal{Q}(\cdot)$ with $B=2$ bits per measurement. The reference voltage for the quantizer was $V_{ref}=0.70\max(\ve{y})$.
    The quantized measurements $\ve{z}=\mathcal{Q}(\ve{y})$ have only $4$ finite values and contain saturations. The total
    transmission bit-budget was only $MB=128$ bits.
\item Fig. \ref{fig:figure01} (c) shows the recovery results of BSBL-BO using real-valued measurements $\ve{y}$.
\item Fig. \ref{fig:figure01} (d)(e)(f) shows the recovery results of QVMP, BSBL-BO and BDQ using the
    quantized measurements $\ve{z}$. QVMP and BDQ used the prior information of quantization cell width $\Delta$ to assist recovery.
\end{enumerate}
Using the same settings in Fig. \ref{fig:figure01}, we presented ARSNR and SSIM of different types of signals on the dataset `subject-02'
in Fig. \ref{fig:figure02}.
The ARSNR was calculated on all $74$ segments in this dataset.
The results in Fig. \ref{fig:figure01} and Fig. \ref{fig:figure02} shows that the proposed algorithm, BDQ,
yielded better signal recoveries from quantized measurements than QVMP and BSBL-BO.

\begin{figure}[!ht]
    \centering
    \includegraphics[width=.99\linewidth]{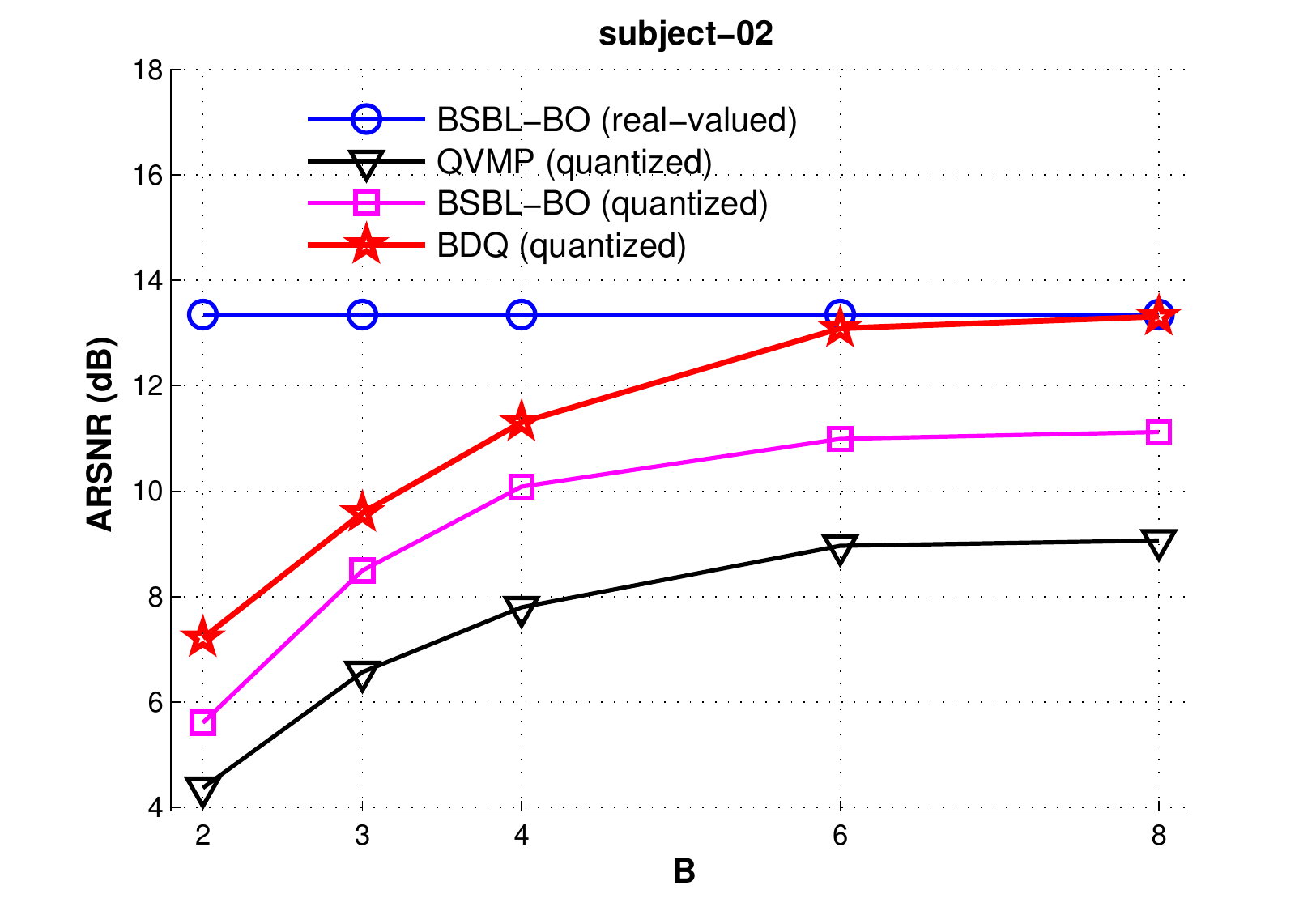}
    \caption{The ARSNR in recovering PPG signals in dataset `subject-02' with multiple quantization bit-depth $B$.}
    \label{fig:figure03}
\end{figure}

We also presented results in recovering PPG signals using multiple quantization bit-depth $B\in\{2,3,4,6,8\}$ in Fig. \ref{fig:figure03}.
The proposed algorithm, BDQ, was in average $1.640$dB superior to BSBL-BO and $3.547$dB to QVMP.
Both BDQ and BSBL-BO were superior to QVMP.

For larger bit-depth $B$ such as $B\in\{6,8\}$,
the variance of the quantization error is small and can be approximated by a Normal distribution.
% explain the discrepancy
However, in Fig. \ref{fig:figure03}, we observed performance gap between BSBL-BO (real-valued) and BSBL-BO (quantized) when
$B\in\{6,8\}$. This was largely caused by the saturation errors of the quantizer,
whose distributions are unbounded and can not be approximated by a Normal distribution.
In contrast, BDQ yielded similar recovery performance to BSBL-BO (real-valued) when $B\in\{6,8\}$.
The reasons were two-fold, one was that the regularization
\eqref{eq:qbsbl_barp_1}-\eqref{eq:qbsbl_barp_2} on the correlation matrix in BDQ can better exploit the highly
correlated structure in PPG signals than the empirical method used in BSBL-BO,
the other was the learning rules \eqref{eq:qbsbl_e} for quantization errors can account for mild saturations introduced by the quantizer.
%Under such circumstances, QVMP was basically a variational Bayesian approach\cite{Tzikas2008} to
%Sparse Bayesian Learning (SBL) \cite{Tipping2001,Wipf2004}.
%The advantage of BSBL-BO to spike sparse algorithms in recovering physiological signals
%has been proved in \cite{Zhang_TBME2012b, liu2013compression}.

%-------- 2 --------
\subsubsection{The trade-off between the compression of CS and the quantization bit-depth}
Quantized CS, when applied to data compression for low-energy telemonitoring, is basically a two-stage compressor.
Firstly, we compress a raw segment of $N$ samples via compressed sensing to achieve a preliminary compression ratio $\mathrm{CR} = (N-M)/N$.
Then the measurements are efficiently encoded by quantization with only $B$ bits per measurement to further reduce the transmission bit-budget.

\begin{figure}[!ht]
    \centering
    \includegraphics[width=\linewidth]{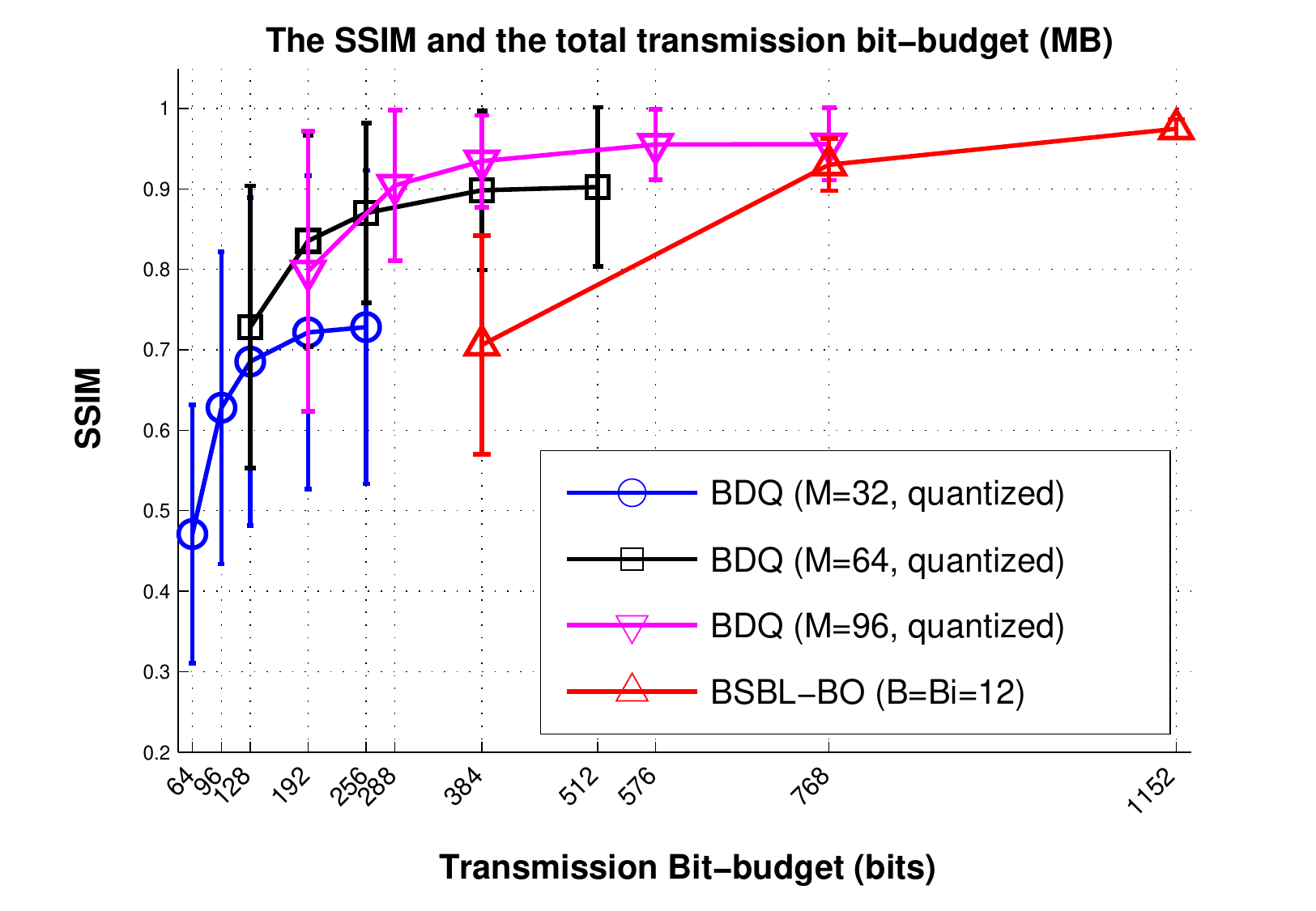}
    \caption{The SSIM w.r.t different transmission bit-budget $MB$ in recovering the PPG channel of dataset `subject-02'.
    We fixed $N=128$ and varied $M\in\{32,64,96\}$. BDQ was used to recover signals from quantized measurements with different
    bit-depth $B$.}
    \label{fig:figure04}
\end{figure}

\begin{figure}[!ht]
    \centering
    \includegraphics[width=\linewidth]{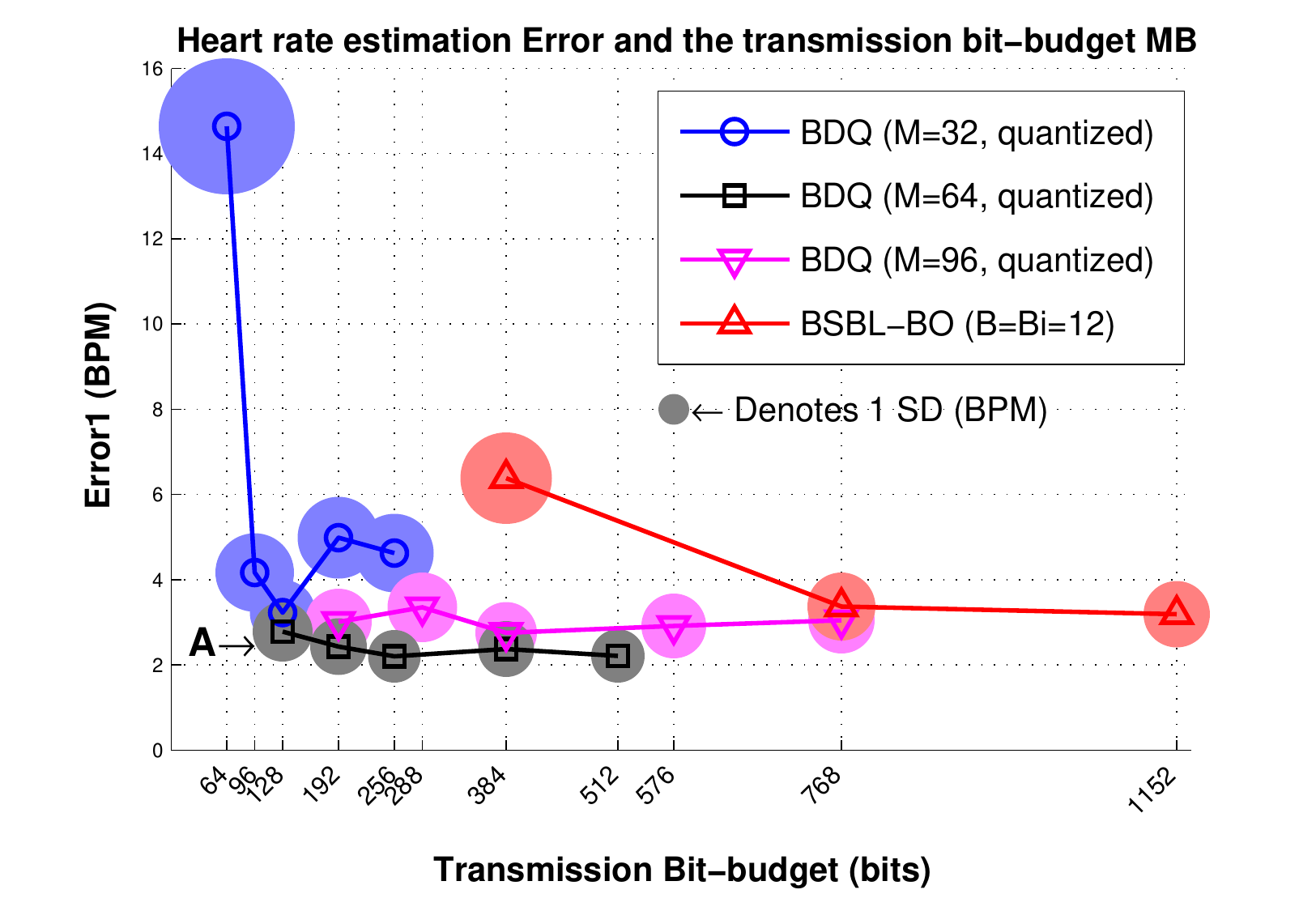}
    \caption{The heart rate estimation error (Error1) and $\mathrm{SD}_{\mathrm{BPM}}$
    w.r.t different transmission bit-budget $MB$. For each configuration $(M,B)$, the diameter of a solid circle is proportional to
    the $\mathrm{SD}_{\mathrm{BPM}}$ metric of this configuration. Point A denotes the optimal configuration where $M=64$ and $B=2$.}
    \label{fig:figure04ext}
\end{figure}

We present in Fig. \ref{fig:figure04} the results with varying number of measurements $M\in\{32,64,96\}$
and quantization bit-depth $B\in\{2,3,4,6,8\}$.
We also present results of traditional \textit{sample-based} compression using BSBL-BO, where the bits per compressed measurement is equal to
the bit-depth of the signals in the dataset, i.e., $B=B_i=12$.
Fig. \ref{fig:figure04ext} shows the absolute heart rate estimation error (Error1) and $\mathrm{SD}_{\mathrm{BPM}}$ with respect to
different transmission bit-budget.

From the results in Fig. \ref{fig:figure04} and Fig. \ref{fig:figure04ext},
we found that both the SSIM and Error1 are affected by the configurations of $M$ and $B$ even if under the same transmission bit-budget.
In fact, CS provides random mixing of signals to compressed measurements, each compressed measurement preserves information on
original signals while the quantization process drops information.
Therefore, if we were going to quantize the compressed measurements with small bit-depth $B$,
the compression ratio $\mathrm{CR}$ can not be high.
In Fig. \ref{fig:figure04ext}, the configuration $M=64$ had the minimal Error1 for $MB < 512$ bits.
It is worth noting that smaller values of $MB$ is always preferable since it reduces total transmission bit-budget.
For $N=128, B_i=12$ and the optimal configuration (point A in Figure. \ref{fig:figure04ext}) of quantized CS $M=64$, $B=2$,
the bit compression ratio $\mathrm{CR}_b$, defined in \eqref{eq:bit_cr}, is
\[
    \mathrm{CR}_b = \frac{NB_i - MB}{NB_i} = 1 - 0.50\cdot\frac{2}{12} = 0.9167.
\]
We achieved $91.67\%$ bit compression ratio and transmitted $128$ bits instead of $128$ samples for telemonitoring.

%-------- 3 --------
\subsubsection{Heart rate estimates from recovered datasets}
%Fig. \ref{fig:figure03} presents the absolute heart rate estimation errors (Error1) from the recovered datasets on `subject-02'
%with different quantization bit-depth $B$.
%%\begin{figure}[!ht]
%%    \centering
%%    \includegraphics[width=2.5in]{figure03}
%%    \caption{The heart rate estimation error with multiple quantization bit-depths.
%%    `un-compressed' denotes the heart rate estimates from uncompressed PPG and accelerometer data.}
%%    \label{fig:figure03}
%%\end{figure}
%The heart rate estimates using the dataset recovered by BDQ was superior and stable.
%The TROIKA framework is basically a frequency based
%method that relies heavily on the correlation informations within a time windowed PPG and accelerometer signals to extract the heart rate.
%The BDQ algorithm imposed KLT regularization to exploit such high correlated information from time series data,
%consequently yielded better signal recovery for TROIKA to estimate the heart rate.

\begin{figure*}[!ht]
    \centering
    \includegraphics[width=0.99\textwidth]{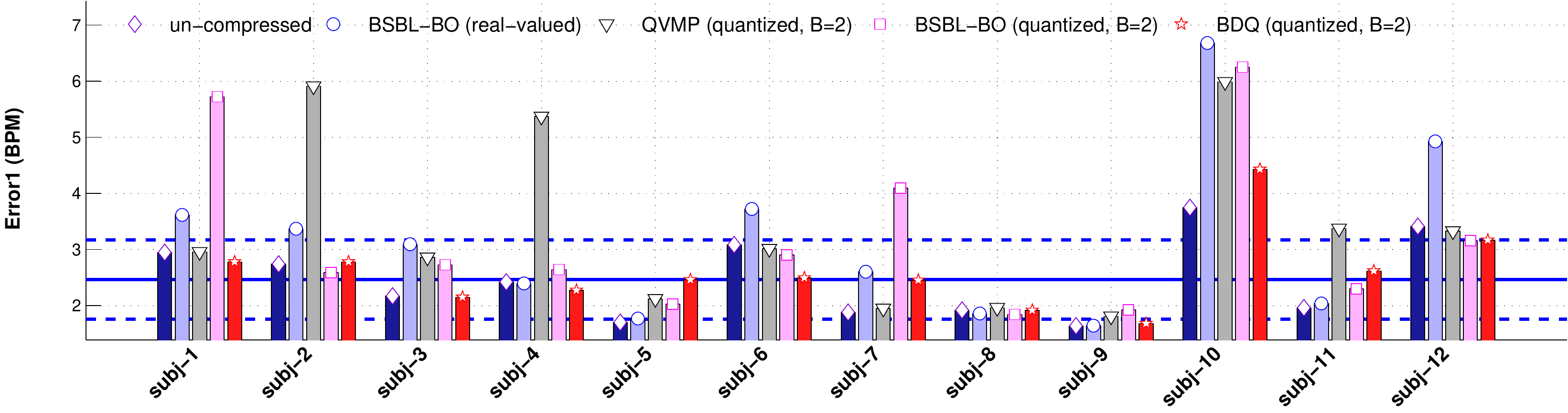}
    \caption{The absolute heart rate estimation error (Error1) on each subject using non-compressed datasets (denoted by `un-compressed')
    and the recovered datasets by different CS algorithms.
    The solid and the dashed line denote mean$\pm$standard deviation of Error1 averaged over all subjects using non-compressed datasets.
    Using the optimal configuration of the quantized CS ($N=128, M=64, \mathrm{CR}=0.50, B=2$) and the proposed BDQ algorithm, we achieved stable and
closely performance to that of `un-compressed'.}
    \label{fig:figure05}
\end{figure*}

We now presented results on the whole datasets using the optimal configurations of quantized CS with $N=128, M=64, \mathrm{CR}=0.50$ and $B=2$.
Fig. \ref{fig:figure05} shows the absolute error (Error1) per subject,
Table \ref{tab:table01} lists absolute error (Error1), Standard Deviation ($\mathrm{SD}_{\mathrm{BPM}}$) and
Pearson correlation calculated over all subjects.
\begin{table*}[!ht]
    \renewcommand{\arraystretch}{1.3}
    \centering
    \caption{The averaged Error1 and the Pearson correlation over all $12$ subjects.}
    \label{tab:table01}
    \begin{tabular}{cccccc}
    \toprule
    & {Uncompressed} & {Real-valued CS} & \multicolumn{3}{c}{Quantized CS ($\mathrm{CR}=0.50, B=2$)} \\
    \cmidrule{4-6}
    &  & {BSBL-BO} & QVMP & BSBL-BO & \textbf{BDQ} \\
    \midrule
    Error1 (BPM) & \textbf{2.464} & 3.137 & 3.355 & 3.179 & \textbf{2.596} \\
    $\mathrm{SD}_{\mathrm{BPM}}$ (BPM) & \textbf{3.554} & 5.022 & 5.561 & 5.030 & \textbf{3.625} \\
    {Pearson Correlation $r$} & \textbf{0.9902} & 0.9810 & 0.9747 & 0.9798 & \textbf{0.9899} \\
    \bottomrule
    \end{tabular}
\end{table*}
By jointly using the optimal configuration for quantized CS and the proposed algorithm, we achieved $Error1 = 2.596$ (BPM),
$\mathrm{SD}_{\mathrm{BPM}}=3.625$ (BPM) and $0.9899$ Pearson correlation, which is closely to the result on
non-compressed datasets. The results in Table. \ref{tab:table01} also shows that for signal recovery from quantized measurements,
BDQ was superior to both QVMP and BSBL-BO.

\begin{figure}[!ht]
    \centering
    \subfloat[BSBL-BO ($\mathrm{CR}=0.50$, $B=2$)]{\includegraphics[width=.49\linewidth]{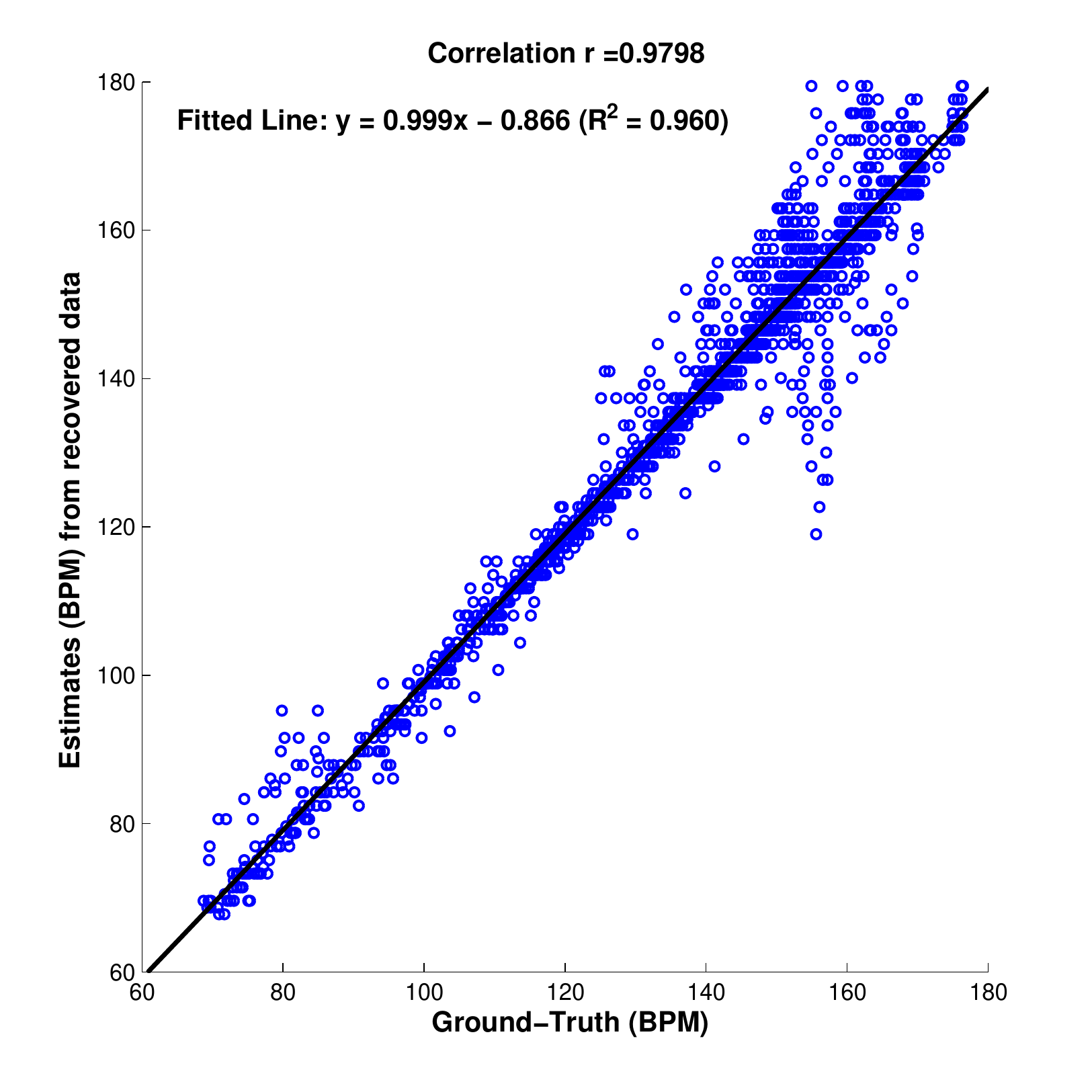}}
    \subfloat[BDQ ($\mathrm{CR}=0.50$, $B=2$)]{\includegraphics[width=.49\linewidth]{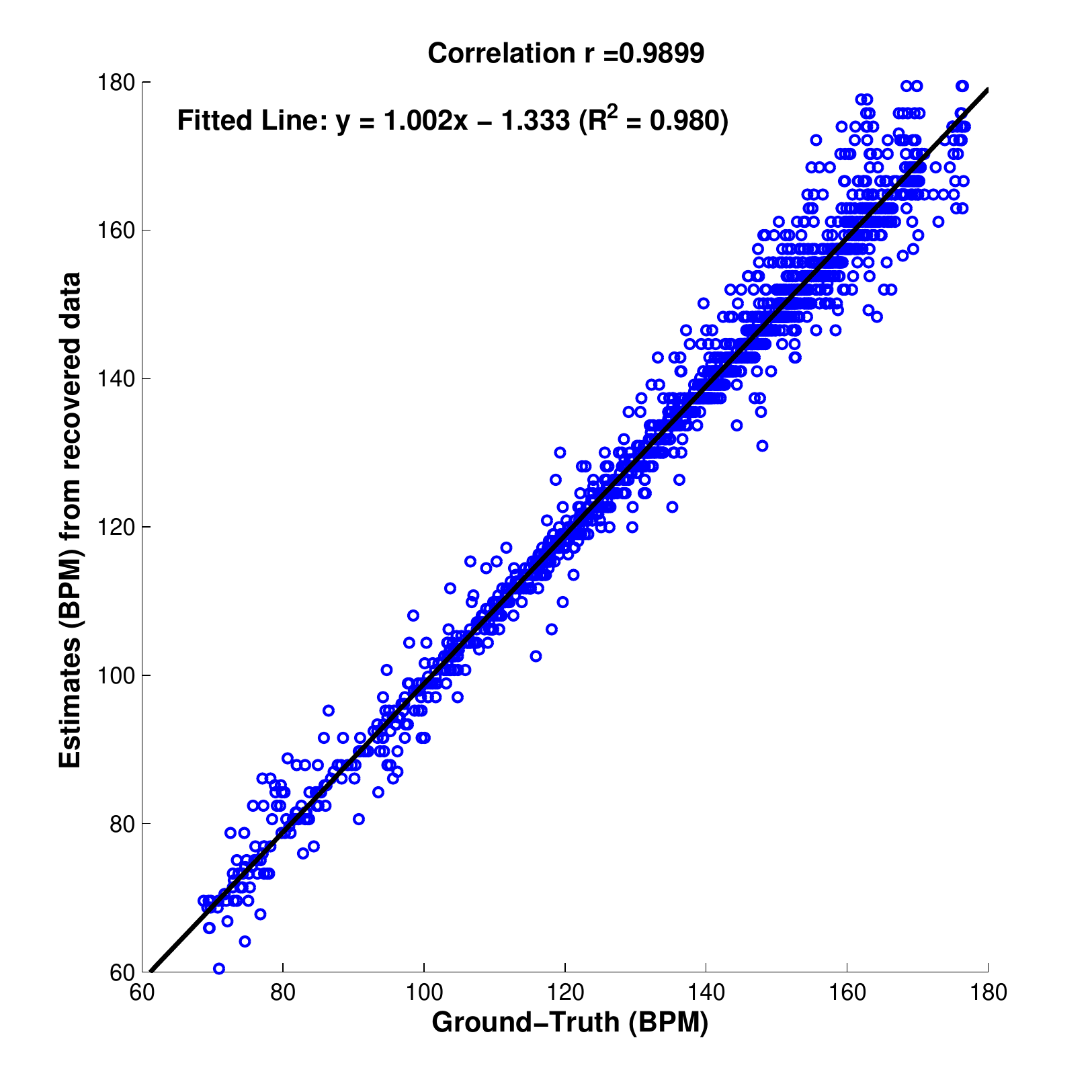}}
    \caption{Scatter Plot between the ground-truth heart rate and the estimates using the recovered data by
        BSBL-BO ($\mathrm{CR}=0.50$, $B=2$) and BDQ ($\mathrm{CR}=0.50$, $B=2$). The Pearson correlation was $0.9798$ and $0.9899$ for
    BSBL-BO and BDQ respectively. The `Fitted Line' is the linear fit of the estimates to ground-truth heart rate.
    The $R^2$ value, which is an estimates for goodness of linear fit, was $0.960$ for BSBL-BO and $0.980$ for BDQ.}
    \label{fig:figure06}
\end{figure}

The Scatter Plot between the ground-truth heart rates and the estimates using the recovered datasets
is shown in Fig. \ref{fig:figure06}.
The fitted line for BSBL-BO ($\mathrm{CR}=0.50$, $B=2$) and BDQ ($\mathrm{CR}=0.50$, $B=2$) was
$y = 0.999x - 0.866, (R^2=0.960)$
and
$y = 1.002x - 1.333, (R^2=0.980)$
respectively, where $x$ indicates the ground-truth heart rate value and $y$ is the estimates
from recovered data, $R^2$ is a measure for goodness of linear fit.

%----
\section{Discussions}

%----
\subsection{The quantizer $\mathcal{Q}(\cdot)$}
The performance of signal recovery from quantized measurements clearly depends on the choice of the quantizer.
For the uniform quantizer used in this paper, the dilemma is the choice of reference voltages $V_{ref}$.

In our experiments, signal $\ve{x}$ was normalized via $\ve{x}/\norm{\ve{x}}$ and then compressed by $\ve{y}=\bm{\Phi}\ve{x}$,
the reference voltage $V_{ref}$ for the quantizer $\mathcal{Q}(\cdot)$ was set to $V_{ref} = 0.70\max(\ve{y})$.
%where $0.70$ was selected arbitrary.
It is worth noting that $V_{ref}=0.70\max(\ve{y})$ is not the optimal reference voltage for this datasets
and we do not search for such optimal values to avoid overfitting.
However one should take care that
for smaller values of $V_{ref}$ there may be more saturations, while for larger values of $V_{ref}$ there may be more underflows.

At the first sight, signal normalization $\ve{x}/\norm{\ve{x}}$ and $\max(\ve{y})$ are not practical for implementing in hardware and
also for low energy applications.
Instead, this problem is solvable by instead fixing the reference voltage for an ADC and using
an Automatic Gain Control (AGC) circuit to tune the scale of signals in between $[-V_{ref},V_{ref}]$.
The gains of AGC must be transmitted alongside with the compressed bits for signal recovery.

%----
\subsection{Signal recovery directly in the time domain}
In literature most CS algorithms recover signals  in a transform domain $\bm{\Psi}$ where signals $\ve{x}$ can be sparsely represented.
By suitably choosing the transformation matrix, one can  improve the quality of recovery.

However, in practice, physiological signals recorded by wearable devices are usually contaminated by various strong artifacts,
such as artifacts due to body motion and hardware issues \cite{Zhang_TBME2012b, Zhang_Asilomar}. As a result, many of these signals
are less sparse in many known transform domains. Seeking/designing an optimal transformation matrix for a specific kind of physiological
signals may be difficult. In this situation, recovering the signals from transform domains is not effective. In \cite{Zhang_TBME2012b},
BSBL-BO was used to recover raw fetal ECG recordings directly in the time domain by exploiting temporal correlations of these recordings,
revealing that exploiting correlations of signals is an alternative method to exploiting sparsity. An obvious advantage of this method
is that it avoids the seeking of an optimal transformation matrix for recovery. Our proposed BDQ algorithm also suggested
the effectiveness of this method.

%----
\subsection{Quantized CS as data encoders for more types of signals}
The CS and the quantizer are analogue to the Discrete Wavelet Transform (DWT) and the scalar quantization unit used
in the JPEG standard\cite{skodras2001jpeg}.
However, CS can be low-energy implemented in FPGA\cite{liu2013energy}
and the additional $2$-bit quantizer barely consumes any resource using fix-point arithmetic\cite{liu2013energy}.
The data compressor proposed in this work can be used as a low-energy data compressor/encoder for potentially more types of signals
such as audios or images.

%----
\section{Conclusions}

In this paper, we present an approach for low-energy wireless telemonitoring using quantized compressed sensing.
The contributions of this paper are two-fold.
First, we propose a two-stage data compressor, where signals are compressed by CS with a compression ratio $\mathrm{CR}=0.50$ and
then quantized with $B=2$ bits per measurement.
Second, to pursue better signal recoveries from quantized measurements, we develop a Bayesian de-quantization algorithm that
can exploit both the model of quantization errors and the correlated structure within signals.
Experiment results showed that, by jointly using the proposed data compressor and the recovery algorithm,
we achieved $2.596$ absolute heart rate estimation errors and $0.9899$ Pearson correlation on whole datasets,
which was closely to the performance on non-compressed datasets.
The results imply that we can effectively transmit $N$ bits instead of $N$ samples, which may revolution the way we compress data for low-energy wireless telemonitoring.

%----
%\section{Discussions}

%\appendices

%% use section* for acknowledgment
%\section*{Acknowledgement}
%The author thanks.

\ifCLASSOPTIONcaptionsoff
  \newpage
\fi

\bibliographystyle{IEEEtran}
\bibliography{bsbl}
%\begin{thebibliography}{1}


%\end{thebibliography}


%\end{IEEEbiography}

%\newpage


%\vfill


\end{document}